\documentclass[%
 reprint,
superscriptaddress,
 amsmath,amssymb,
 aps,
floatfix,nofootinbib,
]{revtex4-2}

\usepackage{graphicx}
\usepackage{dcolumn}
\usepackage{bm}
\usepackage[colorlinks=true,
            linkcolor = blue,
            urlcolor  = blue,
            citecolor = blue,
            anchorcolor = blue]{hyperref}
\usepackage{amsthm}
\usepackage{mathtools}
\usepackage[normalem]{ulem}

\newcommand{\red}{\color{black}}
\newcommand{\Erase}[1]{}

\newcommand{\dd}{\mathrm{d}}

\begin{document}
\title{Thermodynamic role of main reaction pathway and multi-body information flow in membrane transport}
\date{\today}
\author{Satoshi Yoshida}
\affiliation{Department of Physics, Graduate School of Science, the University of Tokyo, 7-3-1 Hongo, Bunkyo-ku, Tokyo 113-0033, Japan}
\author{Yasushi Okada}
\affiliation{Department of Physics, Graduate School of Science, the University of Tokyo, 7-3-1 Hongo, Bunkyo-ku, Tokyo 113-0033, Japan}
\affiliation{Universal Biology Institute (UBI), the University of Tokyo, 7-3-1 Hongo, Bunkyo-ku, Tokyo 113-0033, Japan}
\affiliation{Department of Cell Biology and International Research
Center for Neurointelligence (WPI-IRCN), the University of Tokyo, 7-3-1 Hongo, Bunkyo-ku, Tokyo 113-0033, Japan}
\affiliation{Laboratory for Cell Polarity Regulation, Center for Biosystems Dynamics Research (BDR), RIKEN, 6-2-3 Furue-dai, Suita, Osaka 565-0874, Japan}
\author{Eiro Muneyuki}
\affiliation{Faculty of Science and Engineering, Chuo University, Kasuga, Bunkyo-ku, Tokyo 112-8551, Japan}
\author{Sosuke Ito}
\affiliation{Department of Physics, Graduate School of Science, the University of Tokyo, 7-3-1 Hongo, Bunkyo-ku, Tokyo 113-0033, Japan}
\affiliation{Universal Biology Institute (UBI), the University of Tokyo, 7-3-1 Hongo, Bunkyo-ku, Tokyo 113-0033, Japan}
\affiliation{JST, PRESTO, 4-1-8 Honcho, Kawaguchi, Saitama, 332-0012, Japan}

\begin{abstract}
The two classes of membrane transport, namely, secondary active and passive transport, are understood as different reaction pathways in the same protein structure, \Erase{modeled }{\red described} by the 16-state model in this paper.
To quantify the thermodynamic difference between secondary active transport and passive transport, we extend the second law of information thermodynamics of the autonomous demon in the four-state model composed of two subsystems to the 16-state model composed of four subsystems representing the membrane transport.
We reduce the 16 states to 4 states and derive the coarse-grained second law of information thermodynamics, which provides an upper bound of the free energy transport by the coarse-grained information flow. We also derive an upper bound on the free energy transport by the multi-body information flow representing the two-body or four-body correlations in the 16-state model by exploiting the cycle decomposition. The coarse-grained information flow and the multi-body information flows express the quantitative difference between secondary active and passive transport. The numerical analysis shows that the coarse-grained information flow is positive for secondary active transport and negative for passive transport. The four-body correlation is dominant in the multi-body information flows for secondary active transport. In contrast, the two-body correlation is dominant for passive transport. This result shows that both the sign of the coarse-grained information flow and the difference of the four-body correlation and the two-body correlation explain the difference of the free energy transport in secondary active and passive transport.
\end{abstract}
\maketitle

\section{Introduction}
In living cells, the transport of small molecules across cell membranes is essential to maintain cell homeostasis and communicate information~\cite{roberts2002molecular}. Thermodynamically, small molecules should be transported passively down the concentration gradient without the external driving force. This type of membrane transport without the external driving force is called \emph{passive transport}. A \Erase{membrane transport }protein that transports a single species of substrate (\emph{uniport}) is a simple example of the passive transport. We can also see a different kind of membrane transport against the concentration gradient with the external driving force, called \emph{active transport}, which is also vital for the cell's physiology. The types of external driving forces classify this active transport. Primary active transporters are the protein machinery that typically uses the hydrolysis of adenosine triphosphate (ATP) energy source for transportation. Other transport proteins which use the electrochemical potential generated by this primary active transporter are called \emph{secondary active transporters}. They couple the transport of their substrates with the movement of solutes down their electrochemical potential. Some \Erase{secondary active transporters }{\red of them} transport their substrate in the same direction of the movement of the driver solute (\emph{symport}), while others transport in the opposite direction (\emph{antiport}).
 
Membrane transporters are commonly referred to as solute carrier (SLC) transporter  superfamily, which evolved for the transport of a large variety of different small molecules~\cite{slc,pizzagalli2021guide, SLCreview, SLCreview2}. Traditionally, they have been classified into three distinct classes: uniporters, symporters, and antiporters. Stoichiometry or the coupling between the driver transport and the substrate transport are regarded as tightly fixed. The differences among the three classes of transports (symport, antiport, and uniport) are distinct, and three different mechanisms were assumed. Recent structural studies, however, revealed that SLC proteins share same or similar molecular structures~\cite{bai2017structural}, though their functional roles are scattered among symport, antiport or uniport. Quantitative kinetic measurement of transport also revealed that the coupling between the driver and cargo transports can be variable according to the environmental conditions. In some conditions, the movement of the driver is totally uncoupled from the transport of the substrate. The same protein can show all three modes, antiport, symport and uniport, and the difference between symport, antiport, and uniport is understood as the difference between their main pathways~\cite{bazzone2017loose, MFSantiporter, MFS_sugar_review}. 

Based on these findings, SLC transporter proteins are recently regarded to share the conserved structural mechanisms for transport. The different modes of transportation (symport, antiport, and uniport) would reflect the difference in the main reaction pathways of conformational changes. Intermediate state structures for various SLC proteins have been solved so far, and many molecular dynamics simulation studies have been performed \cite{Forrest2011Structural,Kaback2005Structure, Adelman2016Stochastic, Bisha2016Molecular, Schicker2021Descriptors}. Many mechanical or kinetic models have been proposed for each specific transport mode or transporter protein. For example, a model was recently proposed to model the leaky coupling in a proton-coupled multi-drug transporter protein in bacteria by including all bidirectional transitions between the observed ten states~\cite{Hussey2020Highly}. This model explained various aspects of the experimental results for this specific protein. Still, it is unclear how the qualitative difference of the main reaction pathway makes the quantitative difference in the free energy transport. Thus, a general and unified theoretical framework is awaited to explain the role of the main reaction pathway for free energy transport quantitatively and predictably. 

The framework of stochastic thermodynamics has been developed to describe the thermodynamics of mesoscopic molecular processes~\cite{Sekimoto2010Stochastic,seifert2012stochastic}, which would naturally be applied to the solute transport by the SLC proteins. The recent studies of Maxwell's demon~\cite{Leff1990Maxwell} in stochastic thermodynamics, namely information thermodynamics~\cite{Sagawa2010Generalized, Sagawa2012Fluctuation, Mandal2012Work, Horowitz2010Nonequilibrium, Esposito2011Second, Sagawa2012Fluctuation, Sagawa2012Nonequilibrium, still2012thermodynamics, Parrondo2015Thermodynamics,Goold2016Role}, reveal the thermodynamic role of multi-body correlation between several subsystems. The second law of information thermodynamics for autonomous systems~\cite{allahverdyan2009thermodynamic, Ito2013Information, Horowitz2014Thermodynamics, hartich2014stochastic, horowitz2014second, Ito2015Maxwell, Shiraishi2015Fluctuation, Horowitz2015Multipartite, Rosinberg2016Continuous,  Ito2016Backward, Spinney2016Transfer, Crooks2018Marginal,auconi2019information, Ito2020Unified,wolpert2020uncertainty,nakazato2021geometrical} implies the entropy changes in one subsystem and bathes attached to this subsystem are generally bounded by information flow from other subsystems, which quantifies the multi-body correlation between several subsystems. The results of information thermodynamics can be applied to biological systems~\cite{Ito2013Information, barato2014efficiency,sartori2014thermodynamic, Ito2015Maxwell, Bo2015Thermodynamic,  tenWolde2016Fundamental, Hartich2016Sensory, McGrath2017Biochemical, Boyd2017Correlationpowered, Goldt2017Stochastic, Ouldridge2017Thermodynamics, Loutchko2017Stochastic, Lahiri2017Informationtheoretic, Fang2019Nonequilibrium, Loutchko2020Allosteric} and have been experimentally tested in artificial~\cite{Toyabe2010Experimental, Berut2012Experimental, Koski2014Experimental, Jun2014HighPrecision, Ciliberto2017Experiments, Ribezzi-Crivellari2019Large} and biological systems~\cite{Rico-Pasto2021Dissipation}. 
The transport driven by the interaction between multiple subsystems has been studied, especially in the model of autonomous demon~\cite{strasberg2013thermodynamics, barato2013autonomous, Diana2014Mutual, hartich2014stochastic, Shiraishi2015Role}. {\red The} thermodynamic role of information flow in a cyclic pathway has been discussed~\cite{Horowitz2014Thermodynamics, Yamamoto2016Linear, lin2021nonequilibrium} based on stochastic thermodynamics for cyclic pathways, namely Schnakenberg network theory~\cite{schnakenberg1976network}.
Considering that the coupling between the multiple transport processes, namely driver and cargo, plays an essential role in the secondary active transport, we can introduce the model of autonomous demon for membrane transport and discuss the thermodynamic role of information flow corresponding to the main reaction pathway in the passive and active transport.
{\red Although there are previous works trying to understand membrane transport as Maxwell's demon \cite{mcclare1971chemical, hopfer2002maxwell, flatt2021abc}, this is the first time to properly investigate the quantitative role of information flow in membrane transport using the stochastic thermodynamic model of autonomous demon.}

Information thermodynamics would provide a reasonable way to evaluate a thermodynamic difference between secondary active transport (antiport and symport) and passive transport (uniport). To discuss these differences based on information thermodynamics, we need to introduce two extensions of the discussion for the autonomous demon. One is to extend the $4$-state model~\cite{strasberg2013thermodynamics, barato2013autonomous, Diana2014Mutual, hartich2014stochastic} or the $8$-state model~\cite{Shiraishi2015Role} to the $16$-state model as the model of the autonomous demon because we need to discuss the main pathway in possible $16$ states of SLC transporter proteins which is larger than the observed ten states in the model~\cite{Hussey2020Highly}. To introduce several information flows  which quantify four-body correlation and two-body correlation, the second law of information thermodynamics can be generalized for this $16$-state model. The other extension is to introduce the concept of coarse-graining, where the $16$-state model is reduced to the $4$-state model. For this coarse-grained $4$-state model, we newly obtain the coarse-grained information flow and the coarse-grained second law of information thermodynamics. Based on two new extensions, we discuss thermodynamic role of the cyclic pathway in membrane transport. The free energy transport in membrane transport is generally bounded by the information flow due to the second law of information thermodynamics. Thermodynamic differences between the secondary active and passive transports can be characterized by the coarse-grained information flow in the coarse-grained model and information flow in the $16$-state model. The coarse-grained information flow is negative for the passive transport, which means that the passive transporter does not perform as the autonomous demon. The major contribution of information flow in the $16$-state model is the two-body correlation for this passive transport. On the contrary,  the coarse-grained information flow is positive for the secondary active transport, which means that the secondary active transporter performs as the autonomous demon. The major contribution of information flow in the $16$-state model is the four-body correlation for this secondary active transport. This fact implies that the different modes of transportation in SLC transporter proteins can be characterized by information flow. Our result can lead to a unified framework for membrane transport that can model all possible spectrum of transport modes from the viewpoint of information flow quantitatively. 

This paper is organized as follows.
In Section \ref{sec:secondary_transport_and_uniport}, we state the mechanism of the passive and secondary active transports in a unified way by introducing possible 16 states. In Section \ref{sec:16-state_model_for_second_transport}, we define the $16$-state model and describe the passive and secondary active transports in the 16-state model. We show the cycles of the $16$-state model corresponding to the main reaction pathway of the passive and secondary active transporters.
In Section \ref{sec:demon}, we introduce the coarse-graining of the $16$-state model to reduce the $16$ states to $4$ states. We derive the coarse-grained second law of information thermodynamics, which is analogous to the second law of information thermodynamics for the $4$-state model of the autonomous demon.
In Section \ref{sec:picture_16-state_model}, we choose an appropriate cycle basis of the 16-state model and discuss the second law of information thermodynamics for the steady-state using the cycle basis. In the steady-state, the information flows according to the several cycles quantify the multi-body correlation and correspond to the main reaction pathways in the passive and secondary active transports.
In Section \ref{sec:numerical_analysis}, we numerically show that the coarse-grained information flow and the information flow in the $16$-state model quantifies the difference between the passive and secondary active transports.
In Section \ref{sec:conclusion}, we conclude the main results.
\section{16 states in secondary active transport and passive transport}
\label{sec:secondary_transport_and_uniport}

We here discuss and model the mechanisms of the proton-lactose symporter (LacY) \cite{LacY_Review}, the phosphate-G3P antiporter (GlpT) \cite{GlpTstructure} and the glucose uniporter (GLUT) \cite{mammalianGLUTs} as examples of secondary active transporter and passive transporter. Although their modes of transport are very different, their protein structures are very similar, belonging to the major facilitator superfamily (MFS) fold family \cite{bai2017structural, MFSantiporter, MFS_sugar_review}. Structural analyses by crystallography and cryo-electron microscopy have established that the protein transports the solute by switching among three conformational states. The protein opens toward the outside of the cell for solute binding and releasing. After transiently closing from both sides, the protein then opens to the inside of the cell for the solute exchange with the cytoplasm. The transporter protein would control the ligand binding/release and conformational changes to enable solute transport. As discussed below, we model this process by introducing ratchets. A transport process for one solute can be modeled by one ratchet and two particle baths (inside and outside). Thus, the transport of two solutes by these proteins would be modeled with four particle baths and two ratchets in a unified manner. This model can have 16 internal states since each ratchet can have four states. It should be noted here that this generalized abstract model can be applied not only to the MFS fold family transporters but any molecular machines that transport one or two solutes.

\begin{figure}[tbh]
    \centering
    \includegraphics[width=\linewidth]{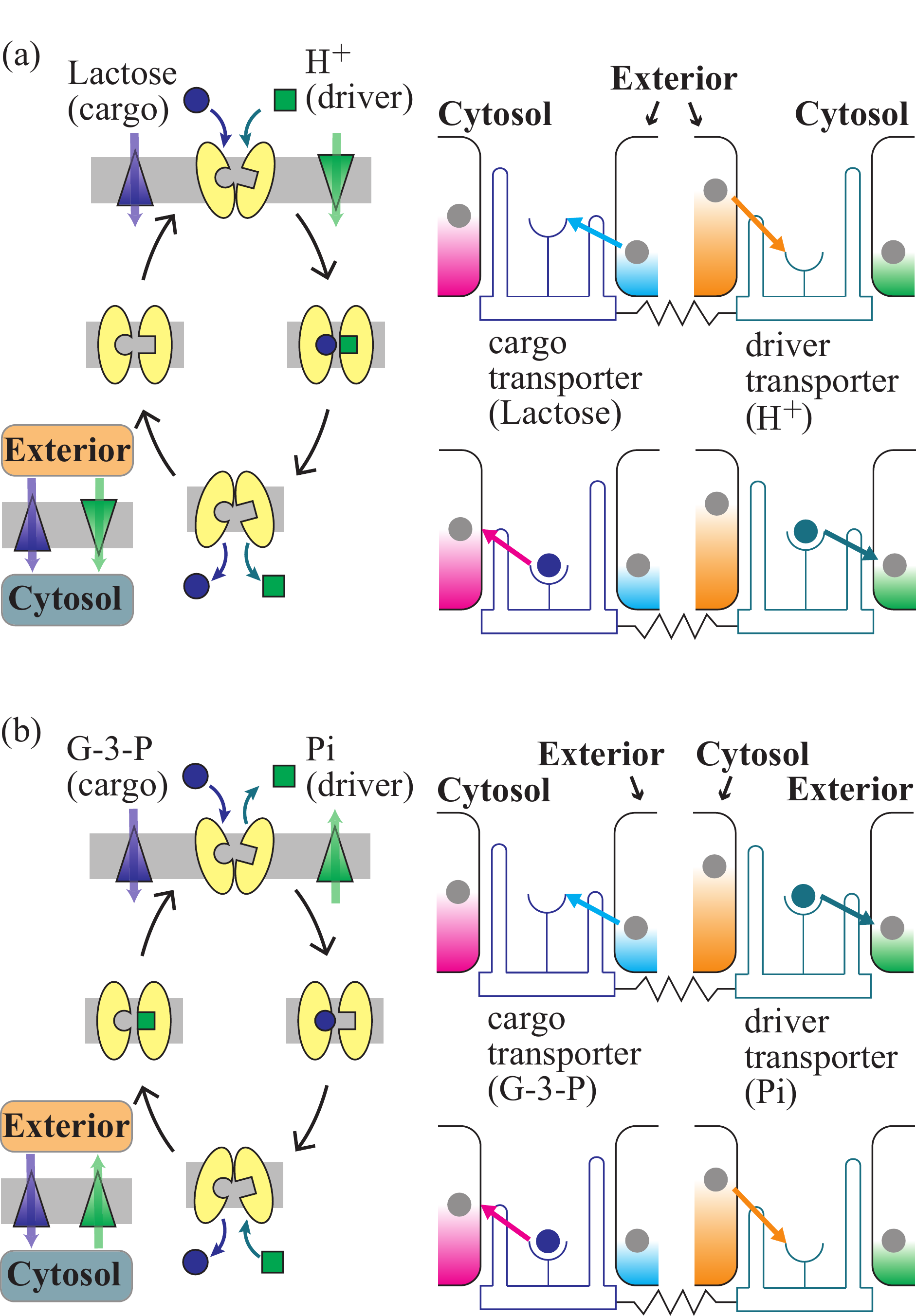}
    \caption{Schematics of the secondary active transporters. (a) Left panel: the reaction pathway of the lactose/proton symporter LacY. The exterior is at a higher proton concentration and a lower lactose concentration than the cytosol. Lactose is transported from the exterior to the cytosol using the free energy generated by the proton transport. Right panel: the symport model of four particle baths and two coupling ratchets. Lactose moves from the exterior to the ratchet when a proton is coupled with the ratchet, and it moves to the cytosol when a proton is released. As a result, one lactose molecule is transported from the exterior to the cytosol. (b) Left panel: the reaction pathway of the glycerol-3-phosphate (G-3-P) transporter GlpT. G-3-P is transported into the cytosol against its concentration gradient by coupling with the transport of the inorganic phosphate (Pi) from the cytosol. Right panel: the antiport model with the same frame as the symport (a).}
    \label{fig:lacy_and_glut}
\end{figure}

\subsection{Symport}

We first discuss the lactose/proton symporter LacY. LacY transports lactose into the cell by coupling the transport of proton as illustrated in FIG.~\ref{fig:lacy_and_glut}~(a). The outside of the plasma membrane (exterior) is kept at a higher proton concentration by proton pump protein (primary active transporter). LacY uses this electrochemical potential for the uptake of lactose into the cell (cytosol). Alternating access model has been proposed to explain this process based on the protein structure \cite{LacY_Review}. First, the protein opens the gate toward the exterior, enabling proton binding. Proton binding triggers to open the second gate to facilitate lactose binding. Lactose binding closes the gates, and the protein now opens the gates toward the cytosol to release proton and lactose. The empty protein returns to the initial outward-open state.

We abstract this mechanism to obtain a general model for secondary active transport \cite{Muneyuki2010Allosteric} made of two coupled ratchets (the right panel of FIG.~\ref{fig:lacy_and_glut}~(a)). We show a list of the essential characteristics of the proton-lactose symporter.

\begin{enumerate}
    \item The exterior and the cytosol have different chemical potentials.
    \item The transporter has binding sites for proton and glucose.
    \item The binding sites for proton and lactose interacts with each other through the conformational changes of the transporter.
    \item Each binding site has two gates toward the exterior and the cytosol.
    \item Open and close of the gates are regulated by the conformational changes of the transporter.
\end{enumerate}

The exterior and the cytosol can be modeled as particle baths with different chemical potentials. We consider two particle baths for each solute molecule (e.g., proton and lactose). Each solute is transported between the two baths. In order to simplify the model without loosing generality, we put a higher concentration compartment on the left side of the model and a lower concentration compartment on the right. Hence, in the case of symport, the driver solute is transported from left to right, and the cargo solute is transported from right to left. 

The transporter can be modeled as coupled ratchets, each of which has one binding site. For simplicity, we assume that there is one binding site for each solute molecule. We represent the open and close of the gates, which regulate the direction of the transport, by the barrier heights between the ratchet and the particle baths. Again, for simplicity, we assume that each ratchet has two conformational states, determining the binding energy and the barrier heights. Finally, to represent the interaction of proton and lactose binding sites through the conformational change, we consider the interaction energy between two ratchets, which depends on the conformational states of the two ratchets.

The system can take possible 16 states in total. For the cargo transporter, lactose in the example of LacY, its state can be expressed by two binaries: the existence of the lactose in the binding site $\{0,1\}$ and the direction of the opening of the gate $\{l,r\}$. The first binary $0$ and $1$ imply the number of the cargo solute (lactose) in the binding site. The second binary $l$ implies that the gate is open to the left bath (cytosol), and $r$ implies opening to the right bath (exterior). The states for the driver transporter (proton) are similarly expressed by two binaries: $\{0,1\}$ and $\{l, r\}$. These binaries have the same meaning (number of solute in the binding site and direction of the opening of the gate). But it should be noted that the higher concentration compartment is put on the left in our definition. Namely, the left bath for proton corresponds to the exterior, and the right bath corresponds to the cytosol (the right panel of FIG.~\ref{fig:lacy_and_glut}~(a)). Thus, the system can take the possible 16 states  $s \in \{0,1\}^2 \times \{l, r \}^2$.

This coupled-ratchet model with 16 possible states can explain the lactose-proton symport as follows. The binding of a proton from the exterior is coupled to the conformational changes in the lactose ratchet to raise the left barrier and to lower the right barrier, which facilitates the binding of lactose from the exterior. The binding of lactose triggers the conformational changes in the proton ratchet to raise the right barrier and to lower the left barrier to enable the proton release to the cytosol. The proton release, again, coupled to the switching of the barrier heights in the lactose ratchet to release the lactose to the cytosol (the right panel of FIG.~\ref{fig:lacy_and_glut}~(a), see Appendix A for a more detailed discussion on the relation to the conventional alternating access model). As a result, one lactose molecule and one proton are transported from the exterior (at a low lactose concentration and a high proton concentration) to the cytosol (at a high lactose concentration and a low proton concentration). Since the transport of the driver and cargo solutes is one-by-one, the chemical potential differences of the solutes should satisfy a certain condition due to the second law of thermodynamics. Suppose $\Delta\mu_a$ and $\Delta\mu_b$ be the cargo and driver transporter's chemical potential differences, respectively. {\red We assume that $\Delta\mu_a$ and $\Delta\mu_b$ are constant.\footnote{\red The concentration gradient of the driver molecules is kept constant by other (primary) active transporters and buffering systems in the cell. The concentration of the cargo molecules is typically in the range of $\mathrm{m mol/L}$ to $\mu \text{mol/L}$, which corresponds to $10^6$ to $10^3$ molecules in a bacterial cell ($\sim 1\;\text{fL}$), and much more outside of the cell. In this work, we discuss stochastic dynamics of a single transport, where a single transport of the cargo molecule does not effectively change the chemical potential for the cargo molecule(s). The time scale of a single transport is assumed to be much faster than the time scale of the change in the concentration gradient.}} Since the free energy of the whole system must decrease, the chemical potentials should satisfy
\begin{align}
    \Delta\mu_a\leq \Delta\mu_b.\label{eq:chemical_potential}
\end{align}
We assume this condition throughout this paper.

\subsection{Antiport}
We next show that the same model can be applied to the antiporters \cite{MFSantiporter}. We take glycerol-3-phosphate (G-3-P) transporter GlpT \cite{GlpTstructure} as an example (FIG.~\ref{fig:lacy_and_glut}~(b)). Here, both the driver solute inorganic phosphate (Pi) and the cargo solute G-3-P are at higher concentrations in the cytosol than the exterior. Therefore, the left and right baths correspond to the cytosol and the exterior for both G-3-P and Pi, respectively. By this setting, the same two coupled ratchets model (16-state model) can be applied to this antiport. The release of Pi from the ratchet to the exterior triggers the conformational changes in the G-3-P ratchet to raise the left barrier and lower the right barrier to facilitate the binding of G-3-P from the exterior to the ratchet. The binding of G-3-P lowers the right barrier and raises the left barrier to enable the binding of Pi from the cytosol. After phosphate binding, the right barrier is raised, and the left barrier is lowered, and G-3-P moves from the ratchet to the cytosol (the right panel of FIG.~\ref{fig:lacy_and_glut}~(b)). As a result, GlpT transports one G-3-P molecule from the exterior (at a low G-3-P concentration) to the cytosol (at a high G-3-P concentration) and one phosphate molecule from the cytosol (at a high phosphate concentration) to the exterior (at a low phosphate concentration). It should be noted here that both symport and antiport are \Erase{modeled }{\red described} similarly in this 16-state model. The driver solute is transported from the left bath to the right bath down the concentration gradient, which is coupled to the leftward transport of the cargo against the gradient. This is because we assigned the left and right baths as the higher and lower concentration compartments, respectively. By this definition, both symport and antiport can be \Erase{modeled }{\red described} in the same 16-state framework (compare (a) and (b) of FIG.~\ref{fig:lacy_and_glut}~).

\subsection{Uniport}
\begin{figure}[tbh]
    \centering
    \includegraphics[width=\linewidth]{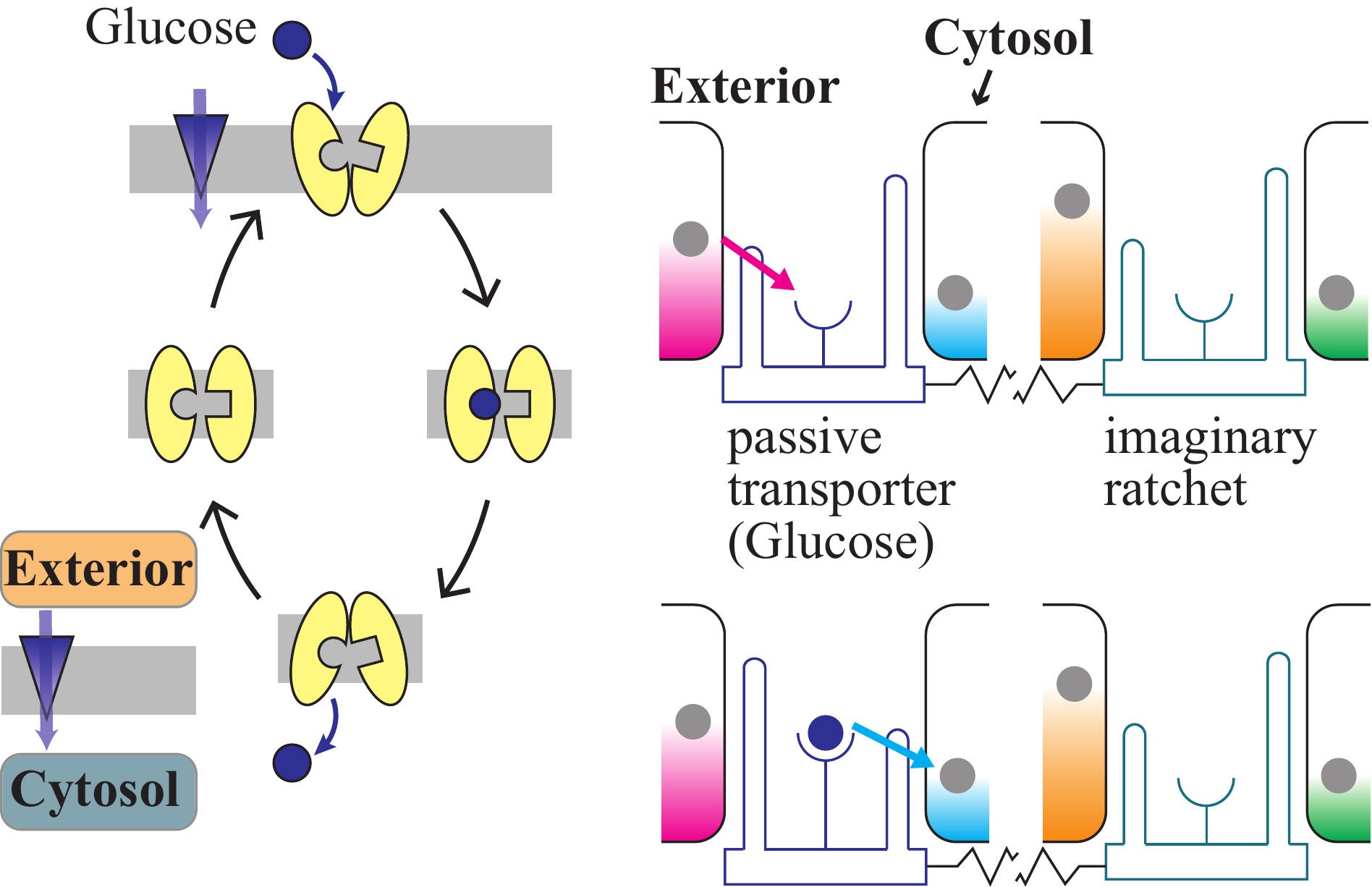}
    \caption{Schematics of passive transporter. Left panel: the reaction pathway of the glucose uniporter GLUT. The exterior is at a higher glucose concentration than the cytosol. Glucose is transported from the exterior to the cytosol down the concentration gradient. Right panel: the same model as the symport with a weak interaction between the ratchets (glucose transporter and imaginary ratchet) can explain this uniport. Since the ratchets are decoupled, glucose transport is driven by the chemical potential difference, and glucose exhibits passive transport.}
    \label{fig:glut}
\end{figure}

Before closing this section, we also show that the same model can be applied to uniport (FIG.~\ref{fig:glut}). For example, the glucose transporter GLUT facilitates the uptake of glucose into the cell down the concentration gradient \cite{MFS_sugar_review, mammalianGLUTs}. Here, the minimal model would need only one ratchet with possible 4 states $\{0,1\} \times \{l, r\}$. Yet, such a minimal model obscures the relation between uniport and symport/antiport. Instead, we would model uniport by the same 16-state model for symport/antiport.  Theoretically, uniport can be \Erase{modeled }{\red described} by the 16-state model by introducing the second imaginary transporter that has two imaginary binary states: $\{0,1\} \times \{l, r\}$. Then, we can discuss the possible 16 states  $s \in \{0,1\}^2 \times \{l, r \}^2$ in parallel with symporter and antiporter. 

Then, uniport can be regarded as the case of a weak or broken interaction between the two ratchets. The two ratchets are not coupled and move independently. The chemical potential difference of cargo solute (glucose) drives the transport. In the case of GLUT uniporter, glucose moves into the ratchet from the left bath (exterior) when the left barrier is down and flows out to the right bath (cytosol) when the right barrier is down. This movement of glucose ratchet is uncoupled from another (imaginary) ratchet. As a result, one glucose molecule moves from the higher concentration compartment (left bath = exterior) to the lower concentration compartment (right bath = cytosol) in one reaction cycle, i.e., glucose exhibits passive transport (FIG.~\ref{fig:glut}).

The introduction of imaginary ratchet might be justified from the evolution of the SLC transport proteins. SLC proteins share similar protein structures and mostly serve as secondary active transporters (symporters or antiporters), but there are many uniporters as well. The bacterial orthologues of GLUT protein are mostly sugar-proton symporters, while mammalian GLUT proteins are mostly uniporters \cite{mammalianGLUTs, MFS_sugar_review}. Interestingly, a bacterial protein GlcP shows an incomplete or loose coupling between proton and sugar transport \cite{bazzone2017loose}.  This wide variety of GLUT protein relatives can be \Erase{modeled }{\red described} by a single 16-state model. The above explained LacY model explains the tightly coupled sugar-proton symport. Loose coupling in GlcP can be explained by weakening the coupling between the sugar transporter and the proton transporter. Then, sugar uniport by GLUT can be understood as the extreme case of the weak coupling. The uncoupled proton channel would have lost its function during evolution \cite{MFS_sugar_review}.

\section{The 16-state model for membrane transport}
\label{sec:16-state_model_for_second_transport}
\subsection{Setup}

\begin{figure}[tbh]
    \centering
    \includegraphics[width=\linewidth]{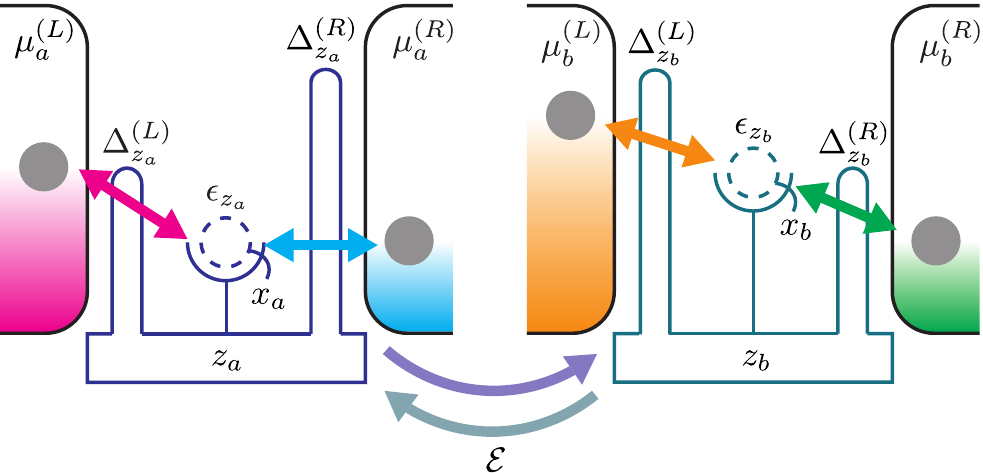}
    \caption{The interaction among the system and the particle baths when the state of the system is $s=(x_a, x_b, z_a, z_b)$. A particle is bound to the ratchet with a bond energy $\epsilon_z$. The height of the barrier between the ratchet and the particle bath $\nu$ is $\Delta_{z}^{(\nu)}$. The interaction energy between the two ratchets is parameterized by $\mathcal{E}$.}
    \label{fig:system}
\end{figure}

We introduce the 16-state model to describe stochastic dynamics of possible 16 states in the membrane transport. The 16-state model is composed of two interacting ratchets, and we assign an index $i \in \{a,b \}$ to the two ratchets (FIG.~\ref{fig:system}). For the secondary active transporter, the ratchets $a$ and $b$ correspond to a cargo transporter and a driver transporter, respectively. For the passive transporter, the ratchet $a$ correspond to the passive transporter and the ratchet $b$ is the pseudo (non-functional) transporter whose function might have been lost in the evolutionary process. Each ratchet is in contact with two particle baths $\nu \in\{L, R\}$ at inverse temperature $\beta$.  The chemical potential of the particle bath $\nu$ attached to the ratchet $i$ is given by $\mu^{(\nu)}_i$. We set that the chemical potential in the particle bath $L$ is larger than the chemical potential in the particle bath $R$, i.e.,
\begin{align}
\mu^{(L)}_i \geq \mu^{(R)}_i.
\end{align}

For the cargo transporter of the LacY, the particle bathes $L$ and $R$ correspond to the cytosol and exterior, respectively. For the driver transporter of the LacY, the particle bathes $L$ and $R$ correspond to the exterior and cytosol, respectively (see also FIG.~\ref{fig:lacy_and_glut}~(a)). For the driver and cargo transporters of the GlpT, the particle bathes $L$ and $R$ correspond to the cytosol and exterior, respectively. (see also FIG.~\ref{fig:lacy_and_glut}~(b)). For the passive transporter of the GLUT, the particle bathes $L$ and $R$ correspond to the exterior and cytosol, respectively (see also FIG.~\ref{fig:glut}). For the secondary imaginary transporter of the GLUT, we also assume that the particle bathes $L$ and $R$ correspond to the exterior and cytosol, respectively.

As shown in FIG.~\ref{fig:system}, we illustrate the particle baths $L$ and $R$ on the left and right sides of each ratchet, respectively. Each ratchet has a site of transporter that can exchange one particle with the attached particle baths. There are barriers between each ratchet and the attached particle baths, and the barrier heights determine the mobility of a particle between the ratchet and the particle baths. The barrier heights correspond to the conformation changes of the transporters.

The state of each ratchet $i$ is represented by two variables $x_i \in \{0,1\}$ and $z_i \in \{l, r\}$. The variable $x_i$ is the number of particles bound to the ratchet $i$. The variable $z_i$ represents the barrier height attached to the ratchet, and $z_i=l$ ($z_i =r$) means that the barrier of the ratchet $i$ against the particle bath $R$ ($L$) is higher than that against the particle bath $L$ ($R$). We call the variable $z_i$ the potential profile of the ratchet $i$. The potential profile $z_i$ fluctuates stochastically by a single heat bath $\nu=L$. The state of the entire system is denoted as $s =(x_a, x_b, z_a, z_b) \in \{ 0,1\}^2 \times \{l,r \}^2 = \mathcal{S}$.

We describe the energetics of the 16-state model. The binding energy of the ratchet $i$ is given by $\epsilon_{z_i}$, i.e., the energy of the ratchet $i$ is $\epsilon_{z_i}$ when a particle is bound to the ratchet, and zero when it is not. The interaction energy of the ratchets $a$ and $b$ is given by $\mathcal{E}\delta_{z_a, z_b}$ using a constant $\mathcal{E}>0$. Here $\delta_{x,y}$ is the Kronecker delta, which is defined by $\delta_{x,x}=1$ and $\delta_{x,y}=0$ for $x\neq y$. In other words, the system is more stable when they have different potential profiles. In total, the energy $E_{x_ax_bz_az_b}$ of the system in the state $s = (x_a,x_b,z_a,z_b)$ is given by
\begin{align}
    E_{x_ax_bz_az_b}=\epsilon_{z_a}x_a+\epsilon_{z_b}x_b+\mathcal{E}\delta_{z_a, z_b}.\label{eq:def_energy}
\end{align}
The height of the barrier between the ratchet $i$ and the particle bath $\nu$ is denoted by $\Delta_{z_i}^{(\nu)}$. The values of $\epsilon_z, \Delta_{z}^{(\nu)}$ are common for the ratchets $a$ and $b$. Because the variable $z_i$ represents the barrier height, the barrier height $\Delta_{z_i}^{(\nu)}$ satisfies the following condition.
\begin{align}
    \Delta_{l}^{(L)}<\Delta_{l}^{(R)}, \; \Delta_{r}^{(R)}<\Delta_{r}^{(L)}.\label{eq:barrier_height}
\end{align}

To achieve the secondary active transport possibly, we assume that the chemical potentials satisfy
\begin{align}
    0<\mu_a^{(L)}-\mu_a^{(R)}<\mu_b^{(L)}-\mu_b^{(R)}, \label{eq:assumption_for_parameter4}
\end{align}
i.e., the chemical potential difference of the cargo transporter $a$ should be smaller than the chemical potential difference of the driver transporter $b$ (see also Eq.~(\ref{eq:chemical_potential})). We also assume the condition
\begin{align}
    \epsilon_l<0<\epsilon_r\label{eq:assumption_for_parameter2}
\end{align}
to induce secondary active transport in the steady state as discussed in subsection \ref{subsec:sec_uni}.

To discuss nonequilibrium dynamics of the membrane transport, we introduce the master equation for stochastic dynamics of the membrane transport. Let $p_{s}$ be the probability of being in state $s$. If the stochastic process is Markovian, the time evolution of the probability distribution $p_s(t)$ is described by the following master equation, 
\begin{align}
    \frac{\dd p_s(t)}{\dd t}=\sum_{\nu, s'}\left[W_{s'\to s}^{(\nu)}p_{s'}-W_{s\to s'}^{(\nu)}p_{s}\right],
\end{align}
where $W_{s\to s'}^{(\nu)}$ represents the transition rate from state $s$ to state $s'$ by bath $\nu$.
We assume the following bipartite condition~\cite{strasberg2013thermodynamics} of the transition rates:
\begin{align}
    W_{s\to s'}^{(\nu)} &= W_{x_a\to x_a'|x_b z_a z_b}^{A (\nu)} \delta_{z_a, z'_a} \delta_{z_b, z'_b}\delta_{x_b, x_b'} (1- \delta_{x_a, x_a'}) \nonumber\\
    &+  W_{x_b\to x_b'|x_a z_a z_b}^{B (\nu)} \delta_{z_a, z'_a} \delta_{z_b, z'_b}\delta_{x_a, x_a'} (1- \delta_{x_b, x_b'}) \nonumber\\
    &+w_{z_a\to z_a'|x_a x_b z_b}^{A} \delta_{x_a, x'_a} \delta_{x_b, x'_b}\delta_{z_b, z_b'} (1- \delta_{z_a, z_a'}) \delta_{\nu, L}\nonumber\\
    &+w_{z_b\to z_b'|x_a x_b z_a}^{B} \delta_{x_a, x'_a} \delta_{x_b, x'_b}\delta_{z_a, z_a'} (1- \delta_{z_b, z_b'}) \delta_{\nu, L},
    \label{eq:bipartite_16-state}
\end{align}
where $s=(x_a,x_b,z_a,z_b)$, $s'=(x_a', x_b', z_a', z_b')$ and $(W^{A(\nu)}, W^{B(\nu)})$ corresponds to the transition rate for the exchange of particles, and $(w^A, w^B)$ corresponds to the transition rate for the change of the potential profile. The bipartite condition means that only one of $X_a$, $X_b$, $Z_a$, and $Z_b$ changes in a single state transition. In stochastic thermodynamics, we assume the local detailed balance conditions~\cite{seifert2012stochastic} to describe the energetics of the stochastic process. For the membrane transport, the local detailed balance conditions of each transition rate $(W^{A(\nu)}, W^{B(\nu)}, w^A, w^B)$ are given by
\begin{align}
    \ln \frac{ W_{0\to 1|x_b z_a z_b}^{A(\nu)} }{  W_{1
    \to 0|x_b z_a z_b}^{A(\nu)}}&= -\beta(E_{1 x_b z_a z_b}-E_{0 x_b z_a z_b}-\mu^{(\nu)}_a),
    \nonumber\\
    \ln \frac{ W_{0\to 1|x_a z_a z_b}^{B(\nu)} }{  W_{1
    \to 0|x_a z_a z_b}^{B(\nu)}}&= -\beta(E_{1 x_b z_a z_b}-E_{0 x_b' z_a z_b}-\mu^{(\nu)}_b),    \nonumber\\
    \ln \frac{ w_{z_a\to z_a'|x_a x_b z_b}^{A}}{  w_{z_a ' \to z_a|x_a x_b z_b}^{A}}&= -\beta(E_{x_a x_b z_a z_b}-E_{x_a x_b z_a' z_b}),
    \nonumber\\
    \ln \frac{ w_{z_b\to z_b'|x_a x_b z_a}^{B}}{  w_{z_b ' \to z_b|x_a x_b z_a}^{B}}&= -\beta(E_{x_a x_b z_a z_b}-E_{x_a x_b z_a z_b'}).\label{eq:detailed balance}
\end{align}
These conditions mean that the dynamics of the 16-state model respects the energetics shown in Eq.~(\ref{eq:def_energy}).
{\red Note that they also implies that all transitions are bidirectional, i.e., $W^{(\nu)}_{s\to s'}\neq 0 \Leftrightarrow W^{(\nu)}_{s'\to s}\neq 0$ for all $s$, $s'$ and $\nu$.} We set the transition rates which satisfy the detailed balance conditions (\ref{eq:detailed balance}) as
\begin{align}
    W^{A(\nu)}_{0\to 1|x_b z_a z_b}&\coloneqq \frac{1}{\tau_0}\exp[-\beta(\Delta^{(\nu)}_{z_a}-\mu^{(\nu)}_a)],\nonumber\\
    W^{A(\nu)}_{1\to 0|x_a z_a z_b}&\coloneqq\frac{1}{\tau_0}\exp[-\beta(\Delta^{(\nu)}_{z_a}-\epsilon_{z_a}],\nonumber\\
    W^{B(\nu)}_{0\to 1|x_a z_a z_b}&\coloneqq\frac{1}{\tau_0}\exp[-\beta(\Delta^{(\nu)}_{z_b}-\mu^{(\nu)}_b)],\nonumber\\
    W^{B(\nu)}_{1\to 0|x_a z_a z_b}&\coloneqq\frac{1}{\tau_0}\exp[-\beta(\Delta^{(\nu)}_{z_b}-\epsilon_{z_b})],\nonumber\\
    w^A_{z_a\to z'_a|x_a x_b z_b}&\coloneqq\frac{1}{\tau_1}\exp[\beta(x_a\epsilon_{z_a}+\mathcal{E}\delta_{z_a, z_b})],\nonumber\\
    w^B_{z_b\to z'_b|x_a x_b z_a}&\coloneqq\frac{1}{\tau_1}\exp[\beta(x_b\epsilon_{z_b}+\mathcal{E}\delta_{z_a, z_b})],\label{eq:def_W}
\end{align}
using the time constants $\tau_0$ and $\tau_1$.

\subsection{The steady state of the 16-state model for secondary active transport and passive transport}
\label{subsec:sec_uni}
\begin{figure*}[tbh]
    \centering
    \includegraphics[width=\linewidth]{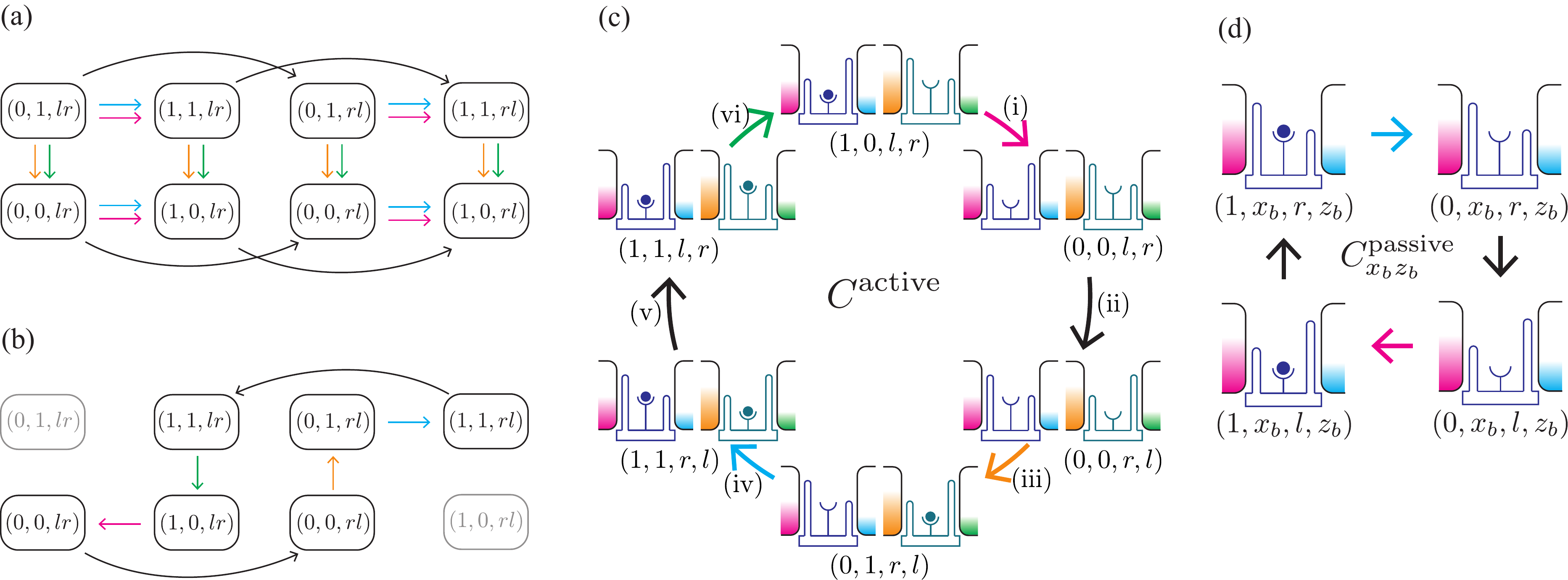}
    \caption{(a) When $\mathcal{E}$ is sufficiently large, two variables $(z_a, z_b)\in \{l,r\}^2$ behave like one variable $z_{ab}\in \{lr, rl\}$ and the 16-state model can be approximated by the 8-state model. The state of the 8-state model is described by $(x_a, x_b, z_{ab})\in \{0,1\}^2\times \{lr,rl\}$ {\red All transitions in the 8-state model are bidirectional. For each transition, either the forward process or the backward process is shown in this figure.} In this graph, each vertex corresponds to a state and each edge corresponds to a transition with the particle bath corresponding to the color of the edge. (b) When the conditions (\ref{eq:assumption_for_parameter4}) and (\ref{eq:assumption_for_parameter2}) holds, the main pathway of the 8-state model becomes the cycle $C'^{\text{active}}$ shown in this figure. (c) The cycle $C'^{\text{active}}$ in the 8-state model corresponds to the cycle $C^{\text{active}}$ in the 16-state model shown in this figure. The cycle $C^{\text{active}}$ corresponds to secondary active transport. Starting from one of the most stable states $s=(1,0,l,r)$, (\romannumeral 1) a particle moves from the ratchet $a$ to the particle bath $L$, (\romannumeral 2) the potential profiles of the ratchets $a$ and $b$ change, (\romannumeral 3) a particle moves from the particle bath $ L$ to the ratchet $b$, (\romannumeral 4) a particles from the particle bath $R$ to the ratchet $a$, (\romannumeral 5) the potential profiles of the ratchets $a, b$ change and (\romannumeral 6) a particle moves from the ratchet $b$ to the particle bath $R$. In each cycle, one particle moves from the particle bath $R$ to $L$ via the ratchet $a$, and one particle moves from the particle bath $L$ to $R$ via the ratchet $b$. (d) When $\mathcal{E}\sim 0$, the main pathway of the 16-state model becomes the cycle $C^{\text{passive}}_{x_b z_b}$ shown in this figure. The cycle $C^{\text{passive}}_{x_b z_b}$ corresponds to passive transport. In each cycle, one particle moves from the particle bath $L$ to $R$ via the ratchet $a$.}
    \label{fig:8-state_graph}
\end{figure*}

We present how secondary active transport and passive transport are described in the steady state of the 16-state model. We denote the probability distribution at steady state of the 16-state model by $p^{\mathrm{ss}}_{s}$. We define the quantity $\mathcal{J}_{a}^{R\to L}$ as
\begin{align}
    \mathcal{J}_{a}^{R\to L}\coloneqq \sum_{x_b, z_a, z_b}& [p^{\mathrm{ss}}_{0 x_b z_a z_b}W^{A(R)}_{0\to 1| x_b z_a z_b} \nonumber\\
    &- p^{\mathrm{ss}}_{1 x_b z_a z_b}W^{A(R)}_{1\to 0| x_b z_a z_b}],\label{eq:J_a^RtoL}
\end{align}
which represents the net transport rate of a particle from the particle bath $R$ to the ratchet $a$. Since the system is in the steady state, the quantity $\mathcal{J}_{a}^{R\to L}$ equals the net transport rate of a particle from the ratchet $a$ to the particle bath $L$. In other words, $\mathcal{J}_{a}^{R\to L}$ is the transport rate of a particle from the particle bath $R$ (for example, the cytosol) to the particle bath $L$ (for example, the exterior) through the ratchet $a$ (for example, the cargo transporter) against the chemical potential difference. In secondary active transport or passive transport, $\mathcal{J}_{a}^{R\to L}$ represents the transport rate of cargo solute through the membrane. The sign of $\mathcal{J}_{a}^{R\to L}$ is positive when the net transport of cargo solute is from the particle bath $R$ to the particle bath $L$. Since $\mu^{(R)}_a<\mu^{(L)}_a$ holds (see Eq.~(\ref{eq:assumption_for_parameter4})), the ratchet $a$ exhibits active transport when $\mathcal{J}_{a}^{R\to L}>0$ and passive transport when $\mathcal{J}_{a}^{R\to L}<0$.
We say that the system exhibits {\it secondary active transport} when
\begin{align}
\mathcal{J}_{a}^{R\to L}>0,
\label{secondaryactivecondition}
\end{align}
and {\it passive transport} when 
\begin{align}
\mathcal{J}_{a}^{R\to L}<0.
\label{passivecondition}
\end{align}

We show the pathway of the 16-state model corresponding to secondary active transport. To this end, we assume that $\mathcal{E}$ is so large that the probability that $z_a=z_b$ is negligible (see Section \ref{sec:numerical_analysis} for a more detailed condition for $\mathcal{E}$ where secondary active transport occurs). In this case, the 16-state model can be approximated by the 8-state model, where the state is described by $(x_a, x_b, z_{ab})\in \{0,1\}^2\times \{lr, rl\}$ and $z_{ab}=lr, rl$ corresponds to $(z_a, z_b)=(l,r), (r,l)$, respectively (FIG.~\ref{fig:8-state_graph}~(a)). {\red In particular, the dominant pathway in the 8-state model is the cycle $C'^{\text{active}}$ shown in  FIG.~\ref{fig:8-state_graph}~(b) (see Appendix~\ref{sec:main_pathway_8-state}).} The pathway $C'^{\text{active}}$ is described as the cycle $C^{\text{active}}$ shown in FIG.~\ref{fig:8-state_graph}~(c) in the 16-state model. See the caption of FIG.~\ref{fig:8-state_graph} for the detail of the cycle $C^{\mathrm{active}}$.

In one cycle of the reaction pathway represented by $C^{\text{active}}$, one particle moves from the particle bath $R$ to $L$ via the ratchet $a$, and one particle moves from the particle bath $L$ to $R$ via the ratchet $b$. Therefore, if only the pathway $C^{\text{active}}$ occurs, the direction of the net transport in the ratchet $a$ is given by
\begin{align}
    \mathcal{J}_{a}^{R\to L}>0.
\end{align}
In this sense, the cycle $C^{\text{active}}$ corresponds to the secondary active transport (see Eq.~(\ref{secondaryactivecondition}) ).

We also show the pathway of the 16-state model corresponding to passive transport. To this end, we consider the case where the ratchets $a$ and $b$ are almost decoupled, i.e., $\mathcal{E}\sim 0$. For the passive transport in the ratchet $a$, dynamics in the ratchet $a$ is not affected by the state of the ratchet $b$, that is ($x_b, z_b$). Then, the dominant pathway for the passive transport is given by the cycle $C^{\mathrm{passive}}_{x_b z_b}$ with the fixed state ($x_b, z_b$) as shown in FIG.~\ref{fig:8-state_graph}~(d). In one cycle of the reaction pathway represented by $C^{\text{passive}}_{x_b z_b}$, one particle moves from the particle bath $L$ to $R$ via the ratchet $a$. 
Therefore, if only the pathway $C^{\text{passive}}_{x_b z_b}$ occurs, the direction of the net transport in the ratchet $a$ is given by
\begin{align}
    \mathcal{J}_{a}^{R\to L}<0.
\end{align}
Similarly to the cycle $C^{\text{active}}$, we can say that the pathway $C^{\mathrm{passive}}_{x_b z_b}$ corresponds to the passive transport (see Eq.~(\ref{passivecondition}) ).

\section{secondary active transport and passive transport in autonomous demon picture}
\label{sec:demon}

In this section, we show how the informational quantity induces the free energy transport in secondary active transport.
To this end, we define the coarse-graining of the 16-state model to reduce 16 states to 4 states, and derive an inequality similar to the second law of information thermodynamics for the autonomous demon in the 4-state model \cite{Horowitz2014Thermodynamics, Yamamoto2016Linear, lin2021nonequilibrium}.

\subsection{Review on the autonomous demon in the 4-state model}

\begin{figure}[tbh]
    \centering
    \includegraphics[width=0.7\linewidth]{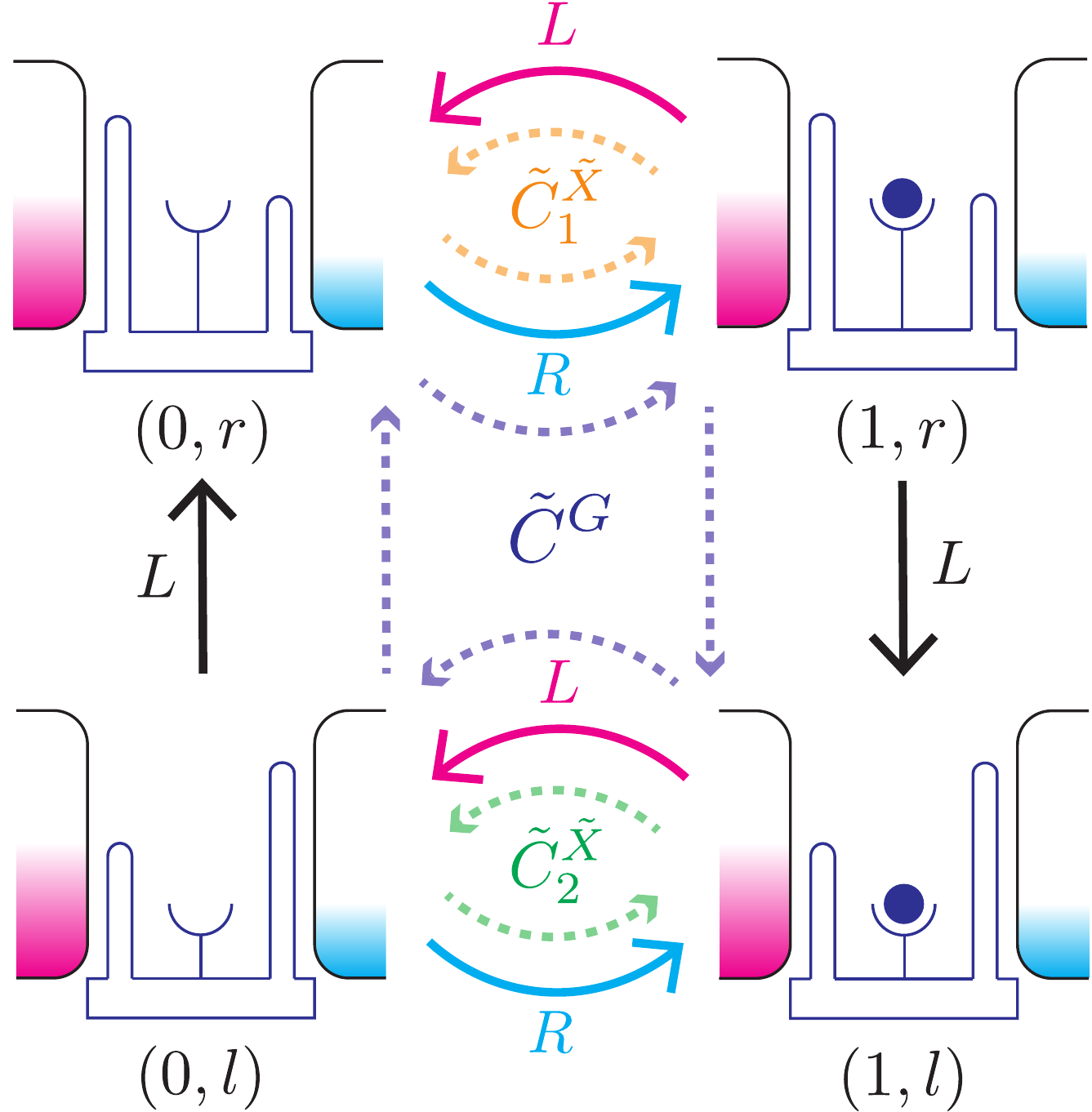}
    \caption{The transitions of the 4-state model are represented by the graph $\tilde{\mathcal{G}}$. The index $L$ or $R$ on each edge corresponds to the particle bath attached when the transition happens.
    The cycles $\tilde{C}^{\tilde{X}}_1$, $\tilde{C}^{\tilde{X}}_2$, and $\tilde{C}^{G}$ form the cycle basis $\tilde{\mathcal{C}}$.}
    \label{fig:4-state}
\end{figure}

We revisit the discussion of the autonomous demon in the 4-state model.
The 4-state model is composed of two coupling subsystems $\tilde{X}$ and $\tilde{Y}$, which represent an engine and a demon, respectively. The states of $\tilde{X}$ and $\tilde{Y}$ are represented as $\tilde{x}\in\{0,1\}$ and $\tilde{y}\in\{l, r\}$, respectively. The system $\tilde{X}$ is in contact with two heat baths $\tilde{\nu}=L, R$ and the system $\tilde{Y}$ is in contact with a single heat bath $\tilde{\nu}=L$. The state of the total system is represented by the random variable $\tilde{S}=(\tilde{X}, \tilde{Y})$, whose realization is $(\tilde{x},\tilde{y})\in \tilde{\mathcal{S}}=\{0,1\}\times \{l,r\}$. We assume that the time evolution of this system is a Markov process. Let $p_{\tilde{s}}$ be the probability of being in state $\tilde{s}$. The probability distribution $p_{\tilde{s}}(t)$ is assumed to evolve according to the master equation given by
\begin{align}
    \frac{\dd p_{\tilde{s}}(t)}{\dd t}=\sum_{\nu, \tilde{s}'}\left[\tilde{W}_{\tilde{s}'\to \tilde{s}}^{(\tilde{\nu})}p_{\tilde{s}'}-\tilde{W}_{\tilde{s}\to \tilde{s}'}^{(\tilde{\nu})}p_{\tilde{s}}\right].
\end{align}
We assume the following bipartite condition for the transition rates $\tilde{W}_{\tilde{s}'\to \tilde{s}}^{(\tilde{\nu})}$:
\begin{align}
    \tilde{W}_{\tilde{s}\to \tilde{s}'}^{(\tilde{\nu})} &= \tilde{W}_{\tilde{x}\to \tilde{x}'|\tilde{y}}^{\tilde{X}(\tilde{\nu})} \delta_{\tilde{y}, \tilde{y}'} (1- \delta_{\tilde{x}, \tilde{x}'}) \nonumber\\
    &+\tilde{W}^{\tilde{Y}}_{\tilde{y}\to \tilde{y}'|\tilde{x}} \delta_{\tilde{x}, \tilde{x}'} (1- \delta_{\tilde{y}, \tilde{y}'})\delta_{\tilde{\nu}, L},
\end{align}
where $\tilde{s}=(\tilde{x},\tilde{y})$ and $\tilde{s}'=(\tilde{x}',\tilde{y}')$.
We also assume that $\tilde{W}_{\tilde{x}\to \tilde{x}'|\tilde{y}}^{\tilde{X}(\tilde{\nu})}$ and $\tilde{W}^{\tilde{Y}}_{\tilde{y}\to \tilde{y}'|\tilde{x}}$ are nonzero. The transitions in the 4-state model can be represented by the graph $\mathcal{\tilde{G}}=(\tilde{\mathcal{V}}, \tilde{\mathcal{E}})$ shown in FIG.~\ref{fig:4-state}, where $\tilde{\mathcal{V}}$ is the set of the vertices and $\tilde{\mathcal{E}}$ is the set of the edges. A vertex $\tilde{s}\in \tilde{\mathcal{V}}$ correspond to the state $\tilde{s}$ and an edge $\tilde{e}=(\tilde{s}\xrightarrow{\tilde{\nu}} \tilde{s}') \in \tilde{\mathcal{E}}$ corresponds to a transition from the state $\tilde{s}$ to the state $\tilde{s}'$ in contact with the bath $\tilde{\nu}$.
We denote the sets of the edges that correspond to the transitions of $\tilde{X}$ and $\tilde{Y}$ by $\tilde{\mathcal{E}}^{\tilde{X}}$ and $\tilde{\mathcal{E}}^{\tilde{Y}}$, respectively.

To show the second law of information thermodynamics, we define the current $J_{\tilde{e}}$, the affinity $F_{\tilde{e}}$, and the effective affinity $\mathcal{F}_{\tilde{e}}$ of an edge $\tilde{e}=(\tilde{s}\xrightarrow{\tilde{\nu}} \tilde{s}')$ as
\begin{align}
    J_{\tilde{e}}&\coloneqq \tilde{W}_{\tilde{s}\to \tilde{s}'}^{(\tilde{\nu})}p_{\tilde{s}}-\tilde{W}_{\tilde{s}'\to \tilde{s}}^{(\tilde{\nu})}p_{\tilde{s}'},\\
    F_{\tilde{e}}&\coloneqq \ln \frac{\tilde{W}^{(\tilde{\nu})}_{\tilde{s}\to\tilde{s}'}}{\tilde{W}^{(\tilde{\nu})}_{\tilde{s}'\to\tilde{s}}},\\
    \mathcal{F}_{\tilde{e}}&\coloneqq \ln \frac{\tilde{W}^{(\tilde{\nu})}_{\tilde{s}\to\tilde{s}'}p_{\tilde{s}}}{\tilde{W}^{(\tilde{\nu})}_{\tilde{s}'\to\tilde{s}}p_{\tilde{s}'}}.
\end{align}
We define $\sigma^{\tilde{X}}$ and $\sigma^{\tilde{Y}}$ as
\begin{align}
    \sigma^{\tilde{X}}&\coloneqq \sum_{\tilde{e}\in \tilde{\mathcal{E}}^{\tilde{X}}}J_{\tilde{e}}\mathcal{F}_{\tilde{e}},\\
    \sigma^{\tilde{Y}}&\coloneqq \sum_{\tilde{e}\in \tilde{\mathcal{E}}^{\tilde{Y}}}J_{\tilde{e}}\mathcal{F}_{\tilde{e}}.
\end{align}
We call $\sigma^{\tilde{X}}$ and $\sigma^{\tilde{Y}}$ the partial entropy productions in $\tilde{X}$ and $\tilde{Y}$, respectively. The partial entropy production is nonnegative because the sign of $J_{\tilde{e}}$ is same as the sign of $\mathcal{F}_{\tilde{e}}$, and $J_{\tilde{e}}\mathcal{F}_{\tilde{e}}\geq 0$ holds for all $\tilde{e}\in\tilde{\mathcal{E}}$.
The nonnegativity of the partial entropy productions is regarded as the second law of information thermodynamics,
\begin{align}
    \sigma^{\tilde{X}}\geq 0,\;\;\;\sigma^{\tilde{Y}}\geq 0.
\end{align}
The second law of information thermodynamics is a generalization of the second law of thermodynamics for a subsystem.

In particular, we consider the steady state, i.e., $\dd p_{\tilde{s}}/\dd t=0$. We denote the probability distribution at the steady state by $p_{\tilde{s}}=p_{\tilde{s}}^{\text{ss}}$. The condition $\dd p_{\tilde{s}}/\dd t=0$ is equivalent to the equation given by
\begin{align}
    \sum_{\tilde{s}', \tilde{\nu}}(J_{\tilde{e}}-J_{\tilde{e}^{\dagger}})=0\;\;\;(\forall \tilde{s}\in \tilde{\mathcal{S}}),\label{eq:stationary_condition}
\end{align}
where $\tilde{e}^{\dagger}\coloneqq (\tilde{s}'\xrightarrow{\tilde{\nu}} \tilde{s})$ is the inverse edge of $e=(\tilde{s}\xrightarrow{\tilde{\nu}} \tilde{s}')$. From the steady state condition (\ref{eq:stationary_condition}), we can introduce the cycle current from the Schnakenberg network theory~\cite{schnakenberg1976network}. To introduce the cycle current, we take a cycle basis $\tilde{\mathcal{C}}=\{\tilde{C_1}, \cdots, \tilde{C}_{\tilde{N}}\}$ of the graph $\tilde{\mathcal{G}}$. We define the cycle matrix $S(\tilde{e},\tilde{C_k})$ for $\tilde{e}\in \tilde{\mathcal{E}}$ and $\tilde{C_k}\in \tilde{\mathcal{C}}$ as
\begin{align}
    S(\tilde{e},\tilde{C_k})\coloneqq
    \begin{cases}
    1 & (\tilde{e}\in \tilde{C_k})\\
    -1 & (\tilde{e}^{\dagger}\in \tilde{C_k})\\
    0 & (\text{otherwise})
    \end{cases}.
    \label{eq:cycle_matrix_definition}
\end{align}
Since Eq.~(\ref{eq:stationary_condition}) is satisfied, the edge current $J_{\tilde{e}}$ can be represented as 
\begin{align}
    J_{\tilde{e}}=\sum_{\tilde{C_k}\in \tilde{\mathcal{C}}} S(\tilde{e}, \tilde{C_k})J(\tilde{C_k})
    \label{eq:cycle_current_definition}
\end{align}
by assigning appropriate values to the cycle currents $J(\tilde{C_k})$ for $\tilde{C_k}\in \tilde{\mathcal{C}}$.
We define the partial affinity $F^{\tilde{X}}(\tilde{C_k})$ and the information affinity $F^I(\tilde{C_k})$ for a cycle $\tilde{C_k}\in \tilde{\mathcal{C}}$ as
\begin{align}
    F^{\tilde{X}}(\tilde{C_k})&\coloneqq \sum_{\tilde{e}\in \tilde{\mathcal{E}}^{\tilde{X}}}S(\tilde{e},\tilde{C_k})F_{\tilde{e}},\\
    F^{I}(\tilde{C_k})&\coloneqq \sum_{\tilde{e}=(\tilde{s}\xrightarrow{\tilde{\nu}}\tilde{s}')\in \tilde{\mathcal{E}}^{\tilde{X}}}S(\tilde{e},\tilde{C_k})\ln \frac{p_{\tilde{s}}^{\text{ss}}}{p_{\tilde{s}'}^{\text{ss}}}.
\end{align}

We here choose the cycle basis $\tilde{\mathcal{C}}=\{\tilde{C}^{\tilde{X}}_1, \tilde{C}^{\tilde{X}}_2, \tilde{C}^{G}\}$ of the graph $\tilde{\mathcal{G}}$
as shown in FIG.~\ref{fig:4-state}. Using this cycle basis, the partial entropy production in {\red $\tilde{X}$} can be written as
\begin{align}
    \sigma^{\tilde{X}}&=\sigma^{\tilde{X}}_{r}+\tilde{\mathcal{I}},\\
    \sigma^{\tilde{X}}_{r}&\coloneqq \sum_{\tilde{C_k}\in\tilde{\mathcal{C}}}J(\tilde{C_k})F^{\tilde{X}}(\tilde{C_k}),\\
    \tilde{\mathcal{I}}&\coloneqq J(\tilde{C}^{G})F^{I}(\tilde{C}^{G}),
\end{align}
where $\sigma^{\tilde{X}}_{r}$ is the change of the entropy of the baths with the change of $\tilde{X}$, and $\tilde{\mathcal{I}}$ is the information flow. {\red The information flow $\tilde{\mathcal{I}}$ is interpreted as a quantity representing how $\tilde{Y}$ measures $\tilde{X}$ \cite{Horowitz2014Thermodynamics}. When $\tilde{\mathcal{I}}>0$, $\tilde{Y}$ gains the information about $\tilde{X}$. When $\tilde{\mathcal{I}}<0$, $\tilde{Y}$ consumes the information about $\tilde{X}$.} 

Defining $\sigma^{\tilde{Y}}_{r}$ in the similar way, we obtain the formula for the partial entropy production in $Y$ given by
\begin{align}
    \sigma^{\tilde{Y}}=\sigma^{\tilde{Y}}_{r}-\tilde{\mathcal{I}}.
\end{align}
Thus, the second law of information thermodynamics is given by
\begin{align}
    \sigma^{\tilde{X}}&=\sigma^{\tilde{X}}_{r}+\tilde{\mathcal{I}}\geq 0,\label{eq:gen_sec_law_4-state}\\
    \sigma^{\tilde{Y}}&=\sigma^{\tilde{Y}}_{r}-\tilde{\mathcal{I}}\geq 0.
\end{align}
The second law of information thermodynamics explains a relation between the entropy change of the subsystem $\sigma^{\tilde{X}}_{r}$ ($\sigma^{\tilde{Y}}_{r}$) and the information flow $\tilde{\mathcal{I}}$.
If $\tilde{\mathcal{I}}>0$, the quantity $\sigma^{\tilde{X}}_r$ can be negative and the second law of thermodynamics in the subsystem $\tilde{X}$ seems to be violated apparently. In other words, the information flow $\tilde{\mathcal{I}}$ compensates the negative entropy change of $\sigma^{\tilde{X}}_r$.
When $\tilde{\mathcal{I}}>0$ and $\sigma^{\tilde{X}}_r<0$, we say that the system $\tilde{X}$ is driven by the autonomous demon $\tilde{Y}$.

\subsection{Coarse-grained picture of the 16-state model}
\label{subsec:coarse-graining}
The driver transporter in the secondary active transport is a possible example of the autonomous demon, because the model of the membrane transport is analogous to the model of the autonomous demon. Thus, we wonder that the importance of the information flow in the secondary active transport can be discussed if we generalize the discussion of the autonomous demon in the 4-state model for the 16-state model of the membrane transport. To discuss it, we here introduce the coarse-grained picture \Erase{of the 16-state model }to reduce 16 states to 4 states, and generalize the statement of the aunotomous demon for the coarse-grained picture.

To discuss the coarse-grained picture of the 16-state model, we regard $(X_a, Z_a)$ as a ``mesostate'' and $(X_b, Z_b)$ as a ``microstate,'' and we focus on the probability distribution of the mesostate $(X_a, Z_a)$~{\red \cite{Esposito2012Stochastica, busiello2020coarse}}. The transitions of the mesostate $(X_a, Z_a)$ can be represented by the graph $\tilde{\mathcal{G}}$ shown in FIG.~\ref{fig:4-state} by identifying $(\tilde{X}, \tilde{Y})$ with $(X_a, Z_a)$. We denote the probability distribution at steady state of the 16-state model by $p_{s}^{\text{ss}}$.

We define the coarse-grained probability distribution $\mathbb{P}_{x_a z_a}^{\text{ss}}$, the coarse-grained edge current $\mathbb{J}_{\tilde{e}}^{\text{ss}}$, and the coarse-grained transition rate $\mathbb{W}^{(\nu)}_{\tilde{s}\to\tilde{s}'}$ of the mesostate at the steady state as
\begin{align}
    \mathbb{P}_{x_a z_a}^{\text{ss}}&\coloneqq \sum_{x_b, z_b}p_{x_ax_bz_az_b}^{\text{ss}},\\
    \mathbb{J}_{\tilde{e}}^{\text{ss}}&\coloneqq \sum_{x_b, z_b}[W^{(\nu)}_{s\to s'}p_{s}^{\text{ss}}-W^{(\nu)}_{s'\to s}p_{s'}^{\text{ss}}],\\
    \mathbb{W}^{(\nu)}_{\tilde{s}\to\tilde{s}'}&\coloneqq \frac{\sum_{x_b, z_b}W^{(\nu)}_{s\to s'}p_{s}^{\text{ss}}}{\mathbb{P}_{\tilde{s}}^{\text{ss}}},
\end{align}
where $s=(x_a, x_b, z_a, z_b)$ and $s'=(x'_a, x_b, z'_a, z_b)$ represents states of the 16-state model, $\tilde{s}=(x_a, z_a)$ and $\tilde{s}'=(x'_a, z'_a)$ represents states of the mesostate, and $\tilde{e}=(\tilde{s}\xrightarrow{\nu}\tilde{s}')$ is an edge of the graph $\widetilde{\mathcal{G}}$.
Since the transition rate $W^{(\nu)}_{s\to s'}$ of the 16-state model satisfies the bipartite condition~(\ref{eq:bipartite_16-state}), the coarse-grained transition rate $\mathbb{W}^{(\nu)}_{\tilde{s}\to\tilde{s}'}$ of the mesostate also satisfies the following bipartite condition:
\begin{align}
    \mathbb{W}_{\tilde{s}\to \tilde{s}'}^{(\nu)} &= \mathbb{W}_{x_a\to x'_a|z_a}^{X_a(\nu)} \delta_{z_a, z'_a} (1- \delta_{x_a, x'_a}) \nonumber\\
    &+\mathbb{W}^{Z_a}_{z_a\to z'_a|x_a} \delta_{x_a, x'_a} (1- \delta_{z_a, z'_a})\delta_{\nu, L},\\
    \mathbb{W}^{X_a (\nu)}_{x_a\to x'_a|z_a}&\coloneqq \frac{\sum_{x_b, z_b}W^{A(\nu)}_{x_a\to x'_a|x_b z_a z_b}p_{s}^{\text{ss}}}{\mathbb{P}_{x_az_a}^{\text{ss}}},\\
    \mathbb{W}^{Z_a}_{z_a\to z'_a|x_a}&\coloneqq \frac{\sum_{x_b, z_b}w^{A}_{z_a\to z'_a|x_a x_b z_b}p_{s}^{\text{ss}}}{\mathbb{P}_{x_az_a}^{\text{ss}}}.
\end{align}
Since the mesostate is in the steady state, the coarse-grained edge current $\mathbb{J}_{\tilde{e}}^{\text{ss}}$ can be represented as
\begin{align}
    \mathbb{J}_{\tilde{e}}^{\text{ss}}=\sum_{\tilde{C_k}\in \tilde{\mathcal{C}}} S(\tilde{e}, \tilde{C_k})\mathbb{J}(\tilde{C_k})
\end{align}
by assigning appropriate values to the coarse-grained cycle currents $\mathbb{J}(\tilde{C_k})$ for $\tilde{C_k}\in \tilde{\mathcal{C}}$.

To show an inequality similar to the second law of information thermodynamics in the 4-state model, we define the coarse-grained partial affinity $\mathbb{F}^{X_a}(\tilde{C_k})$ and the coarse-grained information affinity $\mathbb{F}^I(\tilde{C_k})$ for a cycle $\tilde{C_k}\in \tilde{\mathcal{C}}$ as
\begin{align}
    \mathbb{F}^{X_a}(\tilde{C_k})&\coloneqq \sum_{\tilde{e}\in \tilde{\mathcal{E}}^{\tilde{X}}}S(\tilde{e},\tilde{C_k})\ln \frac{\mathbb{W}_{\tilde{s}\to \tilde{s}'}^{(\nu)}}{\mathbb{W}_{\tilde{s}'\to \tilde{s}}^{(\nu)}},\\
    \mathbb{F}^{I}(\tilde{C_k})&\coloneqq \sum_{\tilde{e}\in \tilde{\mathcal{E}}^{\tilde{X}}}S(\tilde{e},\tilde{C_k})\ln \frac{\mathbb{P}_{\tilde{s}}^{\text{ss}}}{\mathbb{P}_{\tilde{s}'}^{\text{ss}}}.
\end{align}
We define the coarse-grained partial entropy production $\dot{\mathbb{S}}^{X_a}$ in the subsystem $X_a$ as
\begin{align}
    \dot{\mathbb{S}}^{X_a}\coloneqq \sum_{\tilde{e}\in \tilde{\mathcal{E}}^{\tilde{X}}}\mathbb{J}_{\tilde{e}}^{\text{ss}} \ln \frac{\mathbb{W}_{\tilde{s}\to \tilde{s}'}^{(\nu)}\mathbb{P}^{\text{ss}}_{\tilde{s}}}{\mathbb{W}_{\tilde{s}'\to \tilde{s}}^{(\nu)}\mathbb{P}^{\text{ss}}_{\tilde{s}'}}.
\end{align}
Defining the coarse-grained entropy changes $\dot{\mathbb{S}}^{X_a}_{r}$ and the coarse-grained information flow $\mathbb{I}$ by
\begin{align}
    \dot{\mathbb{S}}^{X_a}_{r}&\coloneqq \sum_{\tilde{C_k}\in\tilde{\mathcal{C}}}\mathbb{J}(\tilde{C_k})\mathbb{F}^{X_a}(\tilde{C_k}),\\
    \mathbb{I}&\coloneqq \mathbb{J}(\tilde{C}^{G})\mathbb{F}^{I}(\tilde{C}^{G}),\label{eq:def_I}
\end{align}
we can decompose the coarse-grained partial entropy production in $X_a$ as
\begin{align}
    \dot{\mathbb{S}}^{X_a}&=\dot{\mathbb{S}}^{X_a}_{r}+\mathbb{I}.
\end{align}
Since $W^{A(\nu)}_{x_a\to x'_a|x_b z_a z_b}$ does not depend on $x_b$ nor $z_b$ (see Eq.~(\ref{eq:def_W})),
\begin{align}
    \mathbb{W}^{X_a(\nu)}_{x_a\to x'_a|z_a}=W^{A(\nu)}_{x_a\to x'_a|x_b z_a z_b}
\end{align}
holds. Thus, we obtain
\begin{align}
    \dot{\mathbb{S}}^{X_a}_{r}=\sigma^{X_a}_r,
\end{align}
where $\sigma^{X_a}_r$ is the change of the entropy of the heat baths with the transition of $X_a$ defined by
\begin{align}
    \sigma^{X_a}_r\coloneqq \sum_{e\in\mathcal{E}^{X_a}}J_e^{\text{ss}}\ln\frac{W^{(\nu)}_{s\to s'}}{W^{(\nu)}_{s'\to s}},
\end{align}
where $e=(s\xrightarrow{\nu}s')$ represents the transition in the 16-state model and $\mathcal{E}^{X_a}$ is the set of edges that correspond to the transition of $X_a$. 
Since the coarse-grained edge current $\mathbb{J}_{\tilde{e}}^{\text{ss}}$ can be written as $\mathbb{J}_{\tilde{e}}^{\text{ss}}=\mathbb{W}_{\tilde{s}\to \tilde{s}'}^{(\nu)}\mathbb{P}^{\text{ss}}_{\tilde{s}}-\mathbb{W}_{\tilde{s}'\to \tilde{s}}^{(\nu)}\mathbb{P}^{\text{ss}}_{\tilde{s}'}$,
\begin{align}
    \mathbb{J}_{\tilde{e}}^{\text{ss}} \ln \frac{\mathbb{W}_{\tilde{s}\to \tilde{s}'}^{(\nu)}\mathbb{P}^{\text{ss}}_{\tilde{s}}}{\mathbb{W}_{\tilde{s}'\to \tilde{s}}^{(\nu)}\mathbb{P}^{\text{ss}}_{\tilde{s}'}}\geq 0
\end{align}
holds for all $\tilde{e}\in\tilde{\mathcal{E}}$.
By taking the summation over $\tilde{e}\in\tilde{\mathcal{E}}$, we obtain the coarse-grained second law of information thermodynamics,
\begin{align}
    \dot{\mathbb{S}}^{X_a}=\sigma^{X_a}_{r}+\mathbb{I}\geq 0,\label{eq:gen_second_law}
\end{align}
which is analogous to the second law of information thermodynamics in the 4-state model given by Eq.~(\ref{eq:gen_sec_law_4-state}). When $\mathbb{I} >0$ and $\sigma^{X_a}_{r}<0$, we also say that the system $X_a$ is driven by the autonomous demon $Z_a$. 

We discuss a thermodynamic role of the coarse-grained information flow in the secondary active transport. To discuss it, we show how the transport rate $\mathcal{J}^{R\to L}_{a}$ of cargo solute (see Eq.~(\ref{eq:J_a^RtoL})) appears in the coarse-grained second law of information thermodynamics given by Eq.~(\ref{eq:gen_second_law}).
{\red As shown in Appendix~\ref{sec:relation_transport_rate}, the transport rate $\mathcal{J}^{R\to L}_{a}$ appears in the quantity $\sigma^{X_a}_{r}$ as
\begin{align}
    \sigma^{X_a}_{r}
    =\mathcal{J}^{R\to L}_{a}
    \beta\left(
    \mu^{(R)}_a-\mu^{(L)}_a
    \right)
    +\mathbb{J}(\tilde{C}^{G})
    \beta\left(
    \epsilon_{l}-\epsilon_{r}
    \right).\label{eq:sigma_r}
\end{align}}
Then, the coarse-grained second law of information thermodynamics given by Eq.~(\ref{eq:gen_second_law}) becomes
\begin{align}
    \mathcal{J}^{R\to L}_{a}
    \beta\left(
    \mu^{(R)}_a-\mu^{(L)}_a
    \right)
    +\mathbb{J}(\tilde{C}^{G})
    \beta\left(
    \epsilon_{l}-\epsilon_{r}
    \right)
    +\mathbb{I}\geq 0.
    \label{eq:second_law_X}
\end{align}
This inequality implies that the transport rate $\mathcal{J}_a^{R\to L}$ is bounded by the coarse-grained information flow $\mathbb{I}$. Because the difference between the secondary active transport and the passive transport is characterized by the sign of $\mathcal{J}_a^{R\to L}$, the coarse-grained information flow can quantify the difference between the secondary active transport and the passive transport.

 We discuss a relation between the coarse-grained information flow and the main cyclic pathway of the secondary active transport $C^{\text{active}}$ and the passive transport $C^{\text{passive}}_{x_b z_b}$.
We analyze the behavior of $\sigma_r^{X_a}$ when the main pathway is $C^{\text{active}}$ or $C^{\text{passive}}_{x_b z_b}$ in the 16-state model. The cycle $C^{\text{active}}$ and $C^{\text{passive}}_{x_bz_b}$ corresponds to the cycle $\tilde{C}^G$ and $\tilde{C}^{G\dagger}$ in the coarse-grained picture, respectively (FIG.~\ref{fig:8-state_graph}~(c) and (d) and FIG.~\ref{fig:4-state}). Here, $\tilde{C}^{G\dagger}$ is a cycle whose edges are the same as those of $\tilde{C}^{G}$ but in the opposite direction. Then, if the cycle $C^{\text{active}}$ is dominant in the 16-state model, the dominant pathway in the coarse-grained picture is $\tilde{C}^G$, i.e., $|\mathbb{J}(\tilde{C}^{\tilde{X}}_1)|, |\mathbb{J}(\tilde{C}^{\tilde{X}}_2)|\ll |\mathbb{J}(\tilde{C}^{G})|$ and $\mathbb{J}(\tilde{C}^{G})>0$. In this case, we obtain $\mathcal{J}^{R\to L}_{a}\approx \mathbb{J}(\tilde{C}^G)>0$ that implies the secondary active transport. Since Eqs. (\ref{eq:assumption_for_parameter4}), (\ref{eq:assumption_for_parameter2}) and (\ref{eq:sigma_r}) are satisfied, we obtain
\begin{align}
    \sigma_r^{X_a}<0.\label{eq:sigma_r<0}
\end{align}
Using Eqs. (\ref{eq:second_law_X}) and (\ref{eq:sigma_r<0}), we obtain
\begin{align}
    \mathbb{I}>0.\label{eq:I>0}
\end{align}
When conditions (\ref{eq:sigma_r<0}) and (\ref{eq:I>0}) are satisfied, we can say that the system $X_a$ is driven by the autonomous demon $Z_a$

On the other hand, if the cycle $C^{\text{passive}}_{x_bz_b}$ is dominant in the 16-state model, $\mathcal{J}^{R\to L}_{a}\approx \mathbb{J}(\tilde{C}^G)<0$ holds and thus we obtain
\begin{align}
    \sigma_r^{X_a}>0.
\end{align}
This is the case of the passive transport. In this case, the sign of the quantity $\mathbb{I}$ can be negative. 

As we will see in Section~\ref{sec:numerical_analysis}, $\mathbb{I}>0$ holds when secondary active transport occurs (i.e., $\mathcal{J}_a^{R\to L}>0$) and $\mathbb{I}<0$ holds when the passive transport occurs (i.e., $\mathcal{J}_a^{R\to L}<0$). Therefore, the sign of the coarse-grained information flow $\mathbb{I}$ can distinguish secondary active transport with passive transport.

\section{Multi-body information flow in the 16-state model}
\label{sec:picture_16-state_model}
\begin{figure*}[tb]
    \centering
    \includegraphics[width=\linewidth]{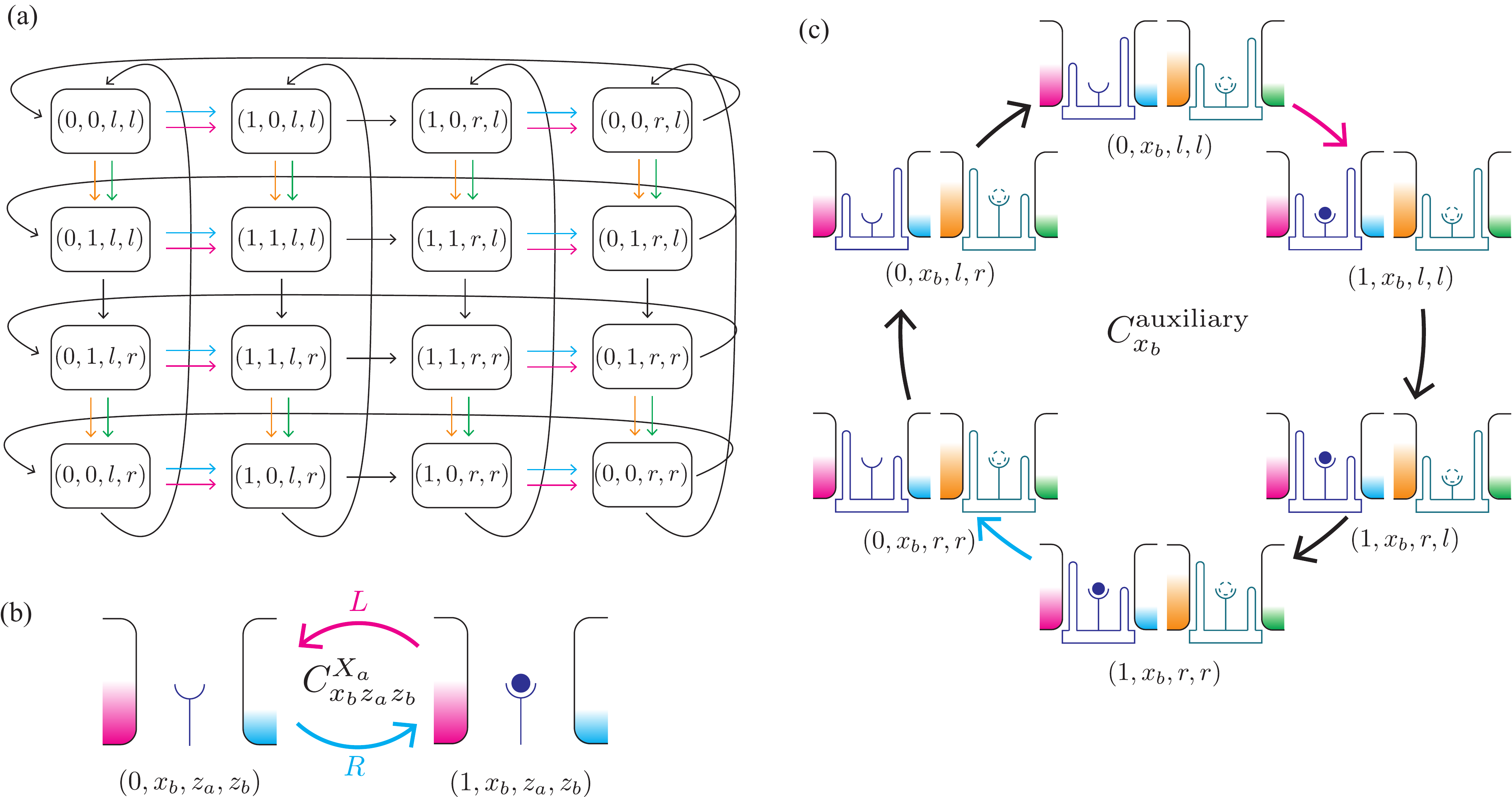}
    \caption{(a) The transitions of the 16-state model are represented by the graph $\mathcal{G}$. Each edge corresponds to a transition with the particle bath corresponding to the color of the edge. (b) The cycle $C^{X_a}_{x_bz_az_b}$ shown in this figure defines the set $\mathcal{C}^{X_a}$ forming a subset of the cycle basis $\mathcal{C}$ (see Eq.~(\ref{eq:C^Xa})). (c) The cycle $C^{\mathrm{auxiliary}}_{x_b}$ shown in this figure is an element of the set $\mathcal{C}^G$ forming a subset of the cycle basis $\mathcal{C}$ (see Eq.~(\ref{eq:C^G})).}
    \label{fig:16-state_graph}
\end{figure*}
In Section \ref{sec:demon}, we discuss the difference between secondary active transport and passive transport in the coarse-grained picture. In this picture, the free energy transport in secondary active transport is induced by the coarse-grained information flow $\mathbb{I}$, and the sign of $\mathbb{I}$ distinguishes secondary active transport with passive transport. However, the coarse-grained information flow $\mathbb{I}$ does not tell which pathway has a dominant correspondence on the partial entropy production in $X_a$. We perform the cycle decomposition \cite{schnakenberg1976network, Andrieux2007Fluctuation} of the graph representing the transitions of the 16-state model to generalize the second law of information thermodynamics and the information flow in the 4-state model to the 16-state model in the steady state. 

We represent the transitions of the 16-state model by the graph $\mathcal{G}=(\mathcal{V}, \mathcal{E})$ shown in FIG.~\ref{fig:16-state_graph}~(a). The set of the vertices is given by $\mathcal{V} = \{ s| s \in \{0,1 \}^2\times \{l,r \}^2 \}$. {\red An edge $e=(s\xrightarrow{\nu} s')$ or its inverse edge $e^\dagger=(s'\xrightarrow{\nu} s)$ is in the set $\mathcal{E}$ if and only if the transition rates $W^{(\nu)}_{s\to s'}$ and $W^{(\nu)}_{s'\to s}$ are nonzero.} We decompose the set of the edges $\mathcal{E}$ into four subsets as
\begin{align}
\mathcal{E}=\mathcal{E}^{X_a}\cup \mathcal{E}^{X_b} \cup \mathcal{E}^{Z_a}\cup \mathcal{E}^{Z_b},
\end{align}
where $\mathcal{E}^{\Omega}$ is the set of edges that correspond to the transitions of the system $\Omega\in \{X_a, X_b, Z_a, Z_b\}$.
Let $\mathcal{C}=\{C_1, C_2, \cdots, C_N\}$ be a cycle basis of the graph $\mathcal{G}$. We define the cycle current $J(C_k)$ and the cycle matrix $S(e, C_k)$ for $e\in \mathcal{E}$ and $C_k\in \mathcal{C}$ similarly to the definitions given by Eqs.~(\ref{eq:cycle_matrix_definition}) and (\ref{eq:cycle_current_definition}) for the 4-state model. We define the partial affinity $F^{X_a}(C_k)$ and the information affinity $F^I(C_k)$ as
\begin{align}
    F^{X_a}(C_k)&\coloneqq \sum_{e\in \mathcal{E}^{X_a}}S(e,C_k)\ln \frac{W^{(\nu)}_{s\to s'}}{W^{(\nu)}_{s'\to s}},\\
    F^{I}(C_k)&\coloneqq \sum_{e\in \mathcal{E}^{X_a}}S(e,C_k)\ln \frac{p_{s}}{p_{s'}},
\end{align}
where $e=(s\xrightarrow{\nu}s')$ and $C_k\in \mathcal{C}$. We decompose the cycle basis into three subsets as $\mathcal{C}=\mathcal{C}^{X_a}\cup \mathcal{C}^{G}\cup \mathcal{C}^{\overline{X_a}}$, where $\mathcal{C}^{X_a}$ is the set of the cycles that are composed only of elements in $\mathcal{E}^{X_a}$, $\mathcal{C}^{G}$ is the set of the cycles that are composed of both elements in $\mathcal{E}^{X_a}$ and those in $\mathcal{E}\setminus \mathcal{E}^{X_a}$, and $\mathcal{C}^{\overline{X_a}}$ is the set of the cycles that do not include elements in $\mathcal{E}^{X_a}$. Then, the partial entropy production in $X_a$ can be written as
\begin{align}
    \sigma^{X_a}&=\sigma^{X_a}_{r}+\mathcal{I},\\
    \sigma^{X_a}_{r}&\coloneqq \sum_{C_k\in\mathcal{C}^{X_a}\cup \mathcal{C}^{G}}J(C_k)F^{X_a}(C_k),\\
    \mathcal{I}&\coloneqq \sum_{C^G_k\in\mathcal{C}^{G}} J(C^G_k)F^{I}(C^G_k),
\end{align}
where $\sigma^{X_a}_{r}$ is the change of the entropy of the baths with the change of $X_a$ and $\mathcal{I}$ is the information flow. {\red As with the information flow $\tilde{\mathcal{I}}$ in the four-state model, $\mathcal{I}$ is interpreted as a quantity representing how the subsystem $X_a$ is measured by the remaining subsystems $X_b, Z_a, Z_b$. When $\mathcal{I}>0$, $X_b, Z_a, Z_b$ gains the information about $X_a$. When $\mathcal{I}<0$, $X_b, Z_a, Z_b$ consumes the information about $X_a$.} The partial entropy production $\sigma^{X_a}$ satisfies the second law of information thermodynamics given by
\begin{align}
    \sigma^{X_a}&=\sigma^{X_a}_{r}+\mathcal{I}\geq 0.\label{eq:gen_sec_law_16-state}
\end{align}

As we discussed in Section~\ref{subsec:coarse-graining}, when the cycle $C^{\mathrm{active}}$ is dominant, $\sigma^{X_a}_r<0$ holds (see Eq.~(\ref{eq:sigma_r<0})). In this case, we obtain $\mathcal{I}>0$ from the second law of information thermodynamics. On the other hand, when the cycle $C^{\mathrm{passive}}_{x_b z_b}$ is dominant, $\sigma^{X_a}_r>0$ and the sign of $\mathcal{I}$ can be negative. We will see that $\mathcal{I}>0$ holds when secondary active transport occurs and $\mathcal{I}<0$ holds when the passive transport occurs from the numerical analysis shown in Section~\ref{sec:numerical_analysis}. Therefore, the information flow $\mathcal{I}$ is useful to distinguish secondary active transport with passive transport.

To see how each cycle in the 16-state model corresponds to the information flow $\mathcal{I}$ and the partial entropy production $\sigma^{X_a}$, we choose a cycle basis of the graph $\mathcal{G}$. The cardinality of $\mathcal{C}$ is given by $N=|\mathcal{E}|-|\mathcal{V}|+1=33$ since $\mathcal{G}$ is a connected graph. We choose the cycle basis such that $|\mathcal{C}^{X_a}|$ and $|\mathcal{C}^{\overline{X_a}}|$ are maximized. In other words, we take $\mathcal{C}^{X_a}$ as a cycle basis of the graph $\mathcal{G}^{X_a}\coloneqq (\mathcal{V}, \mathcal{E}^{X_a})$ and take $\mathcal{C}^{\overline{X_a}}$ as a cycle basis of the graph $\mathcal{G}^{\overline{X_a}}\coloneqq (\mathcal{V}, \mathcal{E}\setminus \mathcal{E}^{X_a})$. Then, $|\mathcal{C}^{X_a}|=8$ and $|\mathcal{C}^{\overline{X_a}}|=18$. In particular, we take $\mathcal{C}^{X_a}$ and $\mathcal{C}^{G}$ as
\begin{align}
    \mathcal{C}^{X_a}=&\{C^{X_a}_{x_b z_a z_b}|x_b\in \{0,1\}, (z_a, z_b)\in \{l,r\}^2\},\label{eq:C^Xa}\\
    \mathcal{C}^G
    =&\{C^{\text{active}}\}\cup \{C^{\text{passive}}_{x_b z_b}|x_b\in\{0,1\}, z_b\in \{l, r\}\}\nonumber\\
    &\cup \{C^{\text{auxiliary}}_{x_b}|x_b\in\{0,1\}\},\label{eq:C^G}
\end{align}
where $C^{X_a}_{x_b z_a z_b}$ is the cycle shown in FIG.~\ref{fig:16-state_graph}~(b), $C^{\text{active}}$ is the cycle corresponding to secondary active transport shown in FIG.~\ref{fig:8-state_graph}~(c), $C^{\text{passive}}_{x_b z_b}$ is the cycle corresponding to passive transport shown in FIG.~\ref{fig:8-state_graph}~(d), and $C^{\text{auxiliary}}_{x_b}$ is an auxiliary cycle shown in FIG.~\ref{fig:16-state_graph}~(c) (see TABLE~\ref{tab:cycle_definition} in Appendix~\ref{sec:cycle_definition}).

Since the quantity $\sigma_r^{X_a}$ is given by Eq.~(\ref{eq:sigma_r}), it can be written using the cycle basis of the 16-state model as
\begin{align}
    \sigma_r^{X_a} 
    =& \mathcal{J}^{R\to L}_{a}
    \beta\left(
    \mu^{(R)}_a-\mu^{(L)}_a
    \right) +J(C^{\mathrm{active}}) \beta\left(
    \epsilon_{l}-\epsilon_{r}
    \right)\nonumber\\
    &-\sum_{x_b} [\sum_{ z_b} J(C^{\mathrm{passive}}_{x_b z_b})+J(C^{\mathrm{auxiliary}}_{x_b})] \beta\left(
    \epsilon_{l}-\epsilon_{r}
    \right).
\end{align}
Then, the second law of information thermodynamics given by Eq. (\ref{eq:gen_sec_law_16-state}) can be written as
\begin{align}
    &\mathcal{J}^{R\to L}_{a}
    \beta\left(
    \mu^{(L)}_a -\mu^{(R)}_a
    \right) \nonumber \\ 
    \leq& \mathcal{I} +J(C^{\mathrm{active}}) \beta\left(
    \epsilon_{l}-\epsilon_{r}
    \right)\nonumber\\
    &  -\sum_{x_b} [\sum_{ z_b} J(C^{\mathrm{passive}}_{x_b z_b})+J(C^{\mathrm{auxiliary}}_{x_b})] \beta\left(
    \epsilon_{l}-\epsilon_{r}
    \right).\label{eq:second_law_16-state_X}
\end{align}
This inequality gives the upper bound of the transport rate $\mathcal{J}_a^{R\to L}$ by the information flow $\mathcal{I}$.

Using the cycle basis of the 16-state model, the information flow $\mathcal{I}$ can be decomposed into three parts
\begin{align}
    \mathcal{I}&=\mathcal{I}^{\text{active}}+\mathcal{I}^{\text{passive}}+\mathcal{I}^{\text{auxiliary}},\\
    \mathcal{I}^{\text{active}}&\coloneqq J(C^{\text{active}})F^{I}(C^{\text{active}}),\\
    \mathcal{I}^{\text{passive}}&\coloneqq 
    \sum_{x_b, z_b}J(C^{\text{passive}}_{x_b z_b})F^{I}(C^{\text{passive}}_{x_b z_b}),\\
    \mathcal{I}^{\text{auxiliary}}&\coloneqq \sum_{x_b}J(C^{\text{auxiliary}}_{x_b})F^{I}(C^{\text{auxiliary}}_{x_b}).
\end{align}
When the cycle $C^{\mathrm{active}}$ corresponding to secondary active transport is dominant, the quantity $\mathcal{I}^{\text{activity}}$ is dominant in the information flow. When the cycle $C^{\mathrm{passive}}_{x_b z_b}$ corresponding to passive transport is dominant, the quantity $\mathcal{I}^{\text{passive}}$ is dominant in the information flow. The quantity $\mathcal{I}^{\text{auxiliary}}$ represents the remaining part of the information flow. We call $\mathcal{I}^{\text{auxiliary}}$ as the auxiliary information flow. As we will confirm from the numerical analysis in Section~\ref{sec:numerical_analysis}, $\mathcal{I}\approx \mathcal{I}^{\text{active}}>0$ holds when secondary active transport occurs. Therefore, the quantity $\mathcal{I}^{\mathrm{active}}$ is the main contribution of the information flow to drive secondary active transport.

To see the informational meaning of the quantities $\mathcal{I}^{\text{active}}$ and $\mathcal{I}^{\text{passive}}$, we define the stochastic multi-information $i(x_a, x_b, z_a, z_b)$ and the stochastic conditional mutual information $i(x_a;z_a|x_b,z_b)$ as
\begin{align}
    i(x_a, x_b, z_a, z_b)&\coloneqq \ln\frac{p^{\text{ss}}_{x_a x_b z_a z_b}}{p^{\text{ss}, X_a}_{x_a}p^{\text{ss}, X_b}_{x_b}p^{\text{ss}, Z_a}_{z_a}p^{\text{ss}, Z_b}_{z_b}},\\
    i(x_a; z_a|x_b, z_b)&\coloneqq \ln\frac{p^{\text{ss}}_{x_a x_b z_a z_b}p^{\text{ss}, X_bZ_b}_{x_b z_b}}{p^{\text{ss}, X_aZ_aZ_b}_{x_az_az_b}p^{\text{ss}, X_bZ_aZ_b}_{x_bz_az_b}},
\end{align}
where $p^{\text{ss},\Omega}$ is a marginal probability distribution of $p^{\mathrm{ss}}$ on the subsystem $\Omega \in \{ X_a, X_b, Z_a, Z_b, X_bZ_b, X_aZ_aZ_b, X_bZ_aZ_b\}$ defined as
\begin{align}
    p^{\text{ss},X_a}_{x_a}&\coloneqq \sum_{x_b, z_a, z_b}p^{\mathrm{ss}}_{x_a x_b z_a z_b},\\
    p^{\text{ss},X_b}_{x_b}&\coloneqq \sum_{x_a, z_a, z_b}p^{\mathrm{ss}}_{x_a x_b z_a z_b},\\
    p^{\text{ss},Z_a}_{z_a}&\coloneqq \sum_{x_a, x_b, z_b}p^{\mathrm{ss}}_{x_a x_b z_a z_b},\\
    p^{\text{ss},Z_b}_{z_b}&\coloneqq \sum_{x_a, x_b, z_a}p^{\mathrm{ss}}_{x_a x_b z_a z_b},\\
    p^{\text{ss},X_b Z_b}_{x_b z_b}&\coloneqq \sum_{x_a, z_a} p^{\mathrm{ss}}_{x_a x_b z_a z_b},\\
    p^{\text{ss},X_a Z_a Z_b}_{x_a z_a z_b}&\coloneqq \sum_{x_b} p^{\mathrm{ss}}_{x_a x_b z_a z_b},\\
    p^{\text{ss},X_b Z_a Z_b}_{x_b z_a z_b}&\coloneqq \sum_{x_a} p^{\mathrm{ss}}_{x_a x_b z_a z_b}.
\end{align}
Note that the expected value of the stochastic multi-information becomes the multi-information (or the total correlation~\cite{watanabe1960information}) given by
\begin{align}
    I[X_a, X_b, Z_a, Z_b]\coloneqq 
    &H(X_a)+H(X_b)+H(Z_a)+H(Z_b)\nonumber\\
    &-H(X_a, X_b, Z_a, Z_b),
\end{align}
where $H(\cdot)$ is the Shannon entropy \cite{Cover1991Elements}. The multi-information is a generalization of the mutual information which quantifies the amount of the multi-body correlation~\cite{Amari2001Information}. 
We remark that the multi-information in stochastic thermodynamics has also been discussed in multiple co-evolving systems~\cite{Wolpert2021Fluctuation}. 

The information affinity is written as a linear combination of the stochastic multi-information and the stochastic conditional mutual information given by
\begin{align}
    F^{I}(C^{\text{active}})=&i(0,1,r,l)+i(1,0,l,r)\nonumber\\
    &-i(1,1,r,l)-i(0,0,l,r),\\
    F^{I}(C^{\text{passive}}_{x_b z_b})=&i(1;r|x_b, z_b)+i(0;l|x_b, z_b)\nonumber\\
    &-i(1;l|x_b, z_b)-i(0;r|x_b, z_b).
\end{align}
Thus, the quantities $\mathcal{I}^{\text{active}}$ and $\mathcal{I}^{\mathrm{passive}}$ are written as
\begin{align}
    \mathcal{I}^{\text{active}}=&J(C^{\mathrm{active}})[i(0,1,r,l)+i(1,0,l,r)\nonumber\\
    &-i(1,1,r,l)-i(0,0,l,r)],\\
    \mathcal{I}^{\mathrm{passive}}=&\sum_{x_b, z_b}J(C^{\mathrm{passive}}_{x_b z_b})[i(1;r|x_b, z_b)+i(0;l|x_b, z_b)\nonumber\\
    &-i(1;l|x_b, z_b)-i(0;r|x_b, z_b)].
\end{align}
In this sense, we call $\mathcal{I}^{\text{active}}$ and $\mathcal{I}^{\text{passive}}$ as {\it multi-body information flows}. The {\red value of} $\mathcal{I}^{\text{active}}$ quantifies the four-body correlation of $X_a$, $X_b$, $Z_a$ and $Z_b$. {\red The value of} $\mathcal{I}^{\text{passive}}$ quantifies the two-body correlation of $X_a$ and $Z_a$ under the condition of {\red $X_b$} and $Z_b$. Since $\mathcal{I}\approx \mathcal{I}^{\mathrm{active}}>0$ holds when secondary active transport occurs, we can say that secondary active transport is driven by the four-body correlation of $X_a$, $X_b$, $Z_a$ and $Z_b$. On the other hand, $\mathcal{I}\approx \mathcal{I}^{\mathrm{passive}}<0$ holds when passive transport occurs, and the passive transport can be characterized by the two-body correlation of $X_a$ and $Z_a$ under the condition of $X_b$ and $Z_b$.  

\section{Numerical analysis}
\label{sec:numerical_analysis}

\begin{figure}[tbh]
    \centering
    \includegraphics[width=\linewidth]{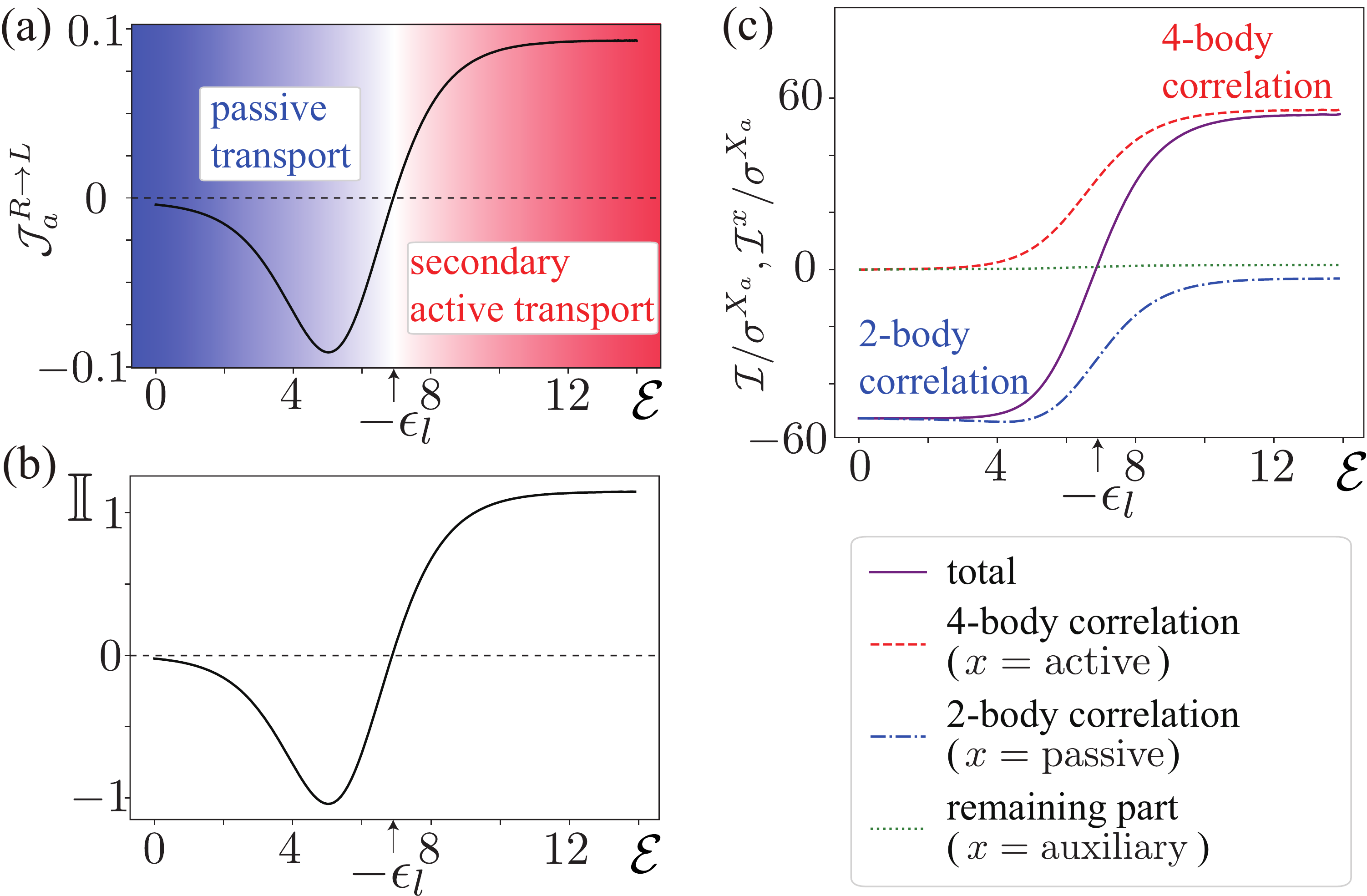}
    \caption{(a) The plot of the transport rate $\mathcal{J}_{a}^{R\to L}$ of cargo solutes through the membrane versus $\mathcal{E}$. When $\mathcal{E}< -\epsilon_l$, $\mathcal{J}_{a}^{R\to L}<0$ holds (i.e., passive transport occurs) and  when $\mathcal{E}> -\epsilon_l$, $\mathcal{J}_{a}^{R\to L}>0$ holds (i.e., secondary active transport occurs).
    (b) The plot of the coarse-grained information flow $\mathbb{I}$ versus $\mathcal{E}$. When passive transport occurs ($\mathcal{E}<-\epsilon_l$), the coarse-grained information flow $\mathbb{I}$ is negative.  When secondary active transport occurs ($\mathcal{E}>-\epsilon_l$), the coarse-grained information flow $\mathbb{I}$ is positive.
    (c) The plot of the information flow $\mathcal{I}$ and the multi-body information flows $\mathcal{I}^{\mathrm{active}}$ and $\mathcal{I}^{\mathrm{passive}}$ and auxiliary information flow $\mathcal{I}^{\mathrm{auxiliary}}$. We plot information flows $\mathcal{I}$ and $\mathcal{I}^x$ for $x\in \{\mathrm{active}, \mathrm{passive}, \mathrm{auxiliary}\}$ normalized by the partial entropy production $\sigma^{X_a}$ versus $\mathcal{E}$. The purple solid line is the normalized total information flow $\mathcal{I}/\sigma^{X_a}$, the red dashed line is the four-body correlation part $\mathcal{I}^{\mathrm{active}}/\sigma^{X_a}$, the blue dashdot line is the two-body correlation part $\mathcal{I}^{\mathrm{passive}}/\sigma^{X_a}$ and the green dotted line is the remaining part $\mathcal{I}^{\mathrm{auxiliary}}/\sigma^{X_a}$. When passive transport occurs ($\mathcal{E}<-\epsilon_l$), the information flow $\mathcal{I}$ is negative and the two-body correlation part is dominant in the information flow. When secondary active transport occurs ($\mathcal{E}>-\epsilon_l$), the information flow $\mathcal{I}$ is positive and the four-body correlation part is dominant in the information flow.}
    \label{fig:plot_flux}
\end{figure}

\begin{figure}[tbh]
    \centering
    \includegraphics[width=\linewidth]{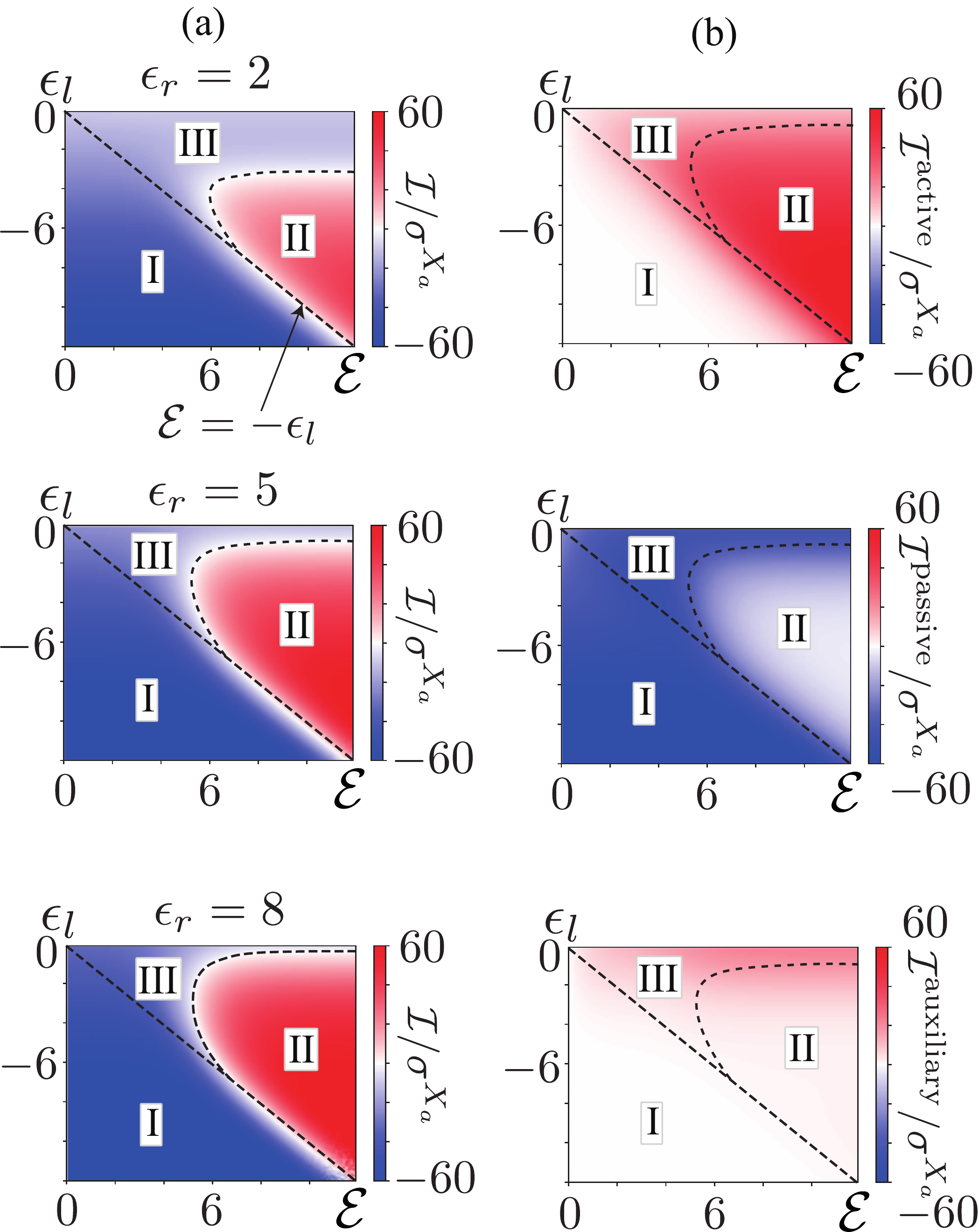}
    \caption{(a) The value of the normalized information flow $\mathcal{I}/\sigma^{X_a}$ for $(\mathcal{E}, \epsilon_l, \epsilon_r) \in [0,12] \times [-12, 0] \times \{2, 5, 8\}$. The $(\mathcal{E}, \epsilon_l)$ plane is divided into three regions I ($\epsilon_l<-\mathcal{E}$), II ($\epsilon_l>\mathcal{E}$ and $\mathcal{I}>0$) and III ($\epsilon_l>\mathcal{E}$ and $\mathcal{I}<0$). Secondary active transport occurs when $\mathcal{E}>-\epsilon_l$ and $-\epsilon_l$ is sufficiently large (region I), but it does not occur even when $\mathcal{E}$ is sufficiently large when $\epsilon_l\sim 0$ (region III). 
    (b) The quantities $\mathcal{I}^{\mathrm{active}}/\sigma^{X_a}$, $\mathcal{I}^{\mathrm{passive}}/\sigma^{X_a}$ and $\mathcal{I}^{\mathrm{auxiliary}}/\sigma^{X_a}$ for $(\mathcal{E}, \epsilon_l)\in [0,12]\times [-12,0]$ and $\epsilon_r=5$ are plotted. In the region I, $\mathcal{I}^{\mathrm{passive}}/\sigma^{X_a}$ is dominant and in the region II, the cycle $\mathcal{I}^{\mathrm{active}}/\sigma^{X_a}$ is dominant. 
    This behavior is consistent with the discussion in Section~\ref{subsec:sec_uni}.
    However, in the region III, not only the quantities $\mathcal{I}^{\mathrm{active}}/\sigma^{X_a}$ and $\mathcal{I}^{\mathrm{passive}}/\sigma^{X_a}$ but also the quantity $\mathcal{I}^{\mathrm{auxiliary}}/\sigma^{X_a}$ correspond to a certain amount of $\mathcal{I}/\sigma^{X_a}$. In other words, the cycles $C^{\mathrm{active}}$ nor $C^{\mathrm{passive}}_{x_b z_b}$ do not represent the main pathway of the 16-state model in the region III.}
    \label{fig:heatmap}
\end{figure}

In this section, we perform numerical calculation to demonstrate that the coarse-grained information flow $\mathbb{I}$ defined in Eq.~(\ref{eq:def_I}), the information flow $\mathcal{I}$, and the multi-body information flows $\mathcal{I}^{\mathrm{passive}}$ and $\mathcal{I}^{\mathrm{active}}$ distinguish secondary active transport with passive transport (see Appendix \ref{appendix:numerical_analysis_detail} for the numerical condition). To this end, we calculate the transport rate $\mathcal{J}_{a}^{R\to L}$ of cargo solutes, the quantity $\mathbb{I}$, the information flow $\mathcal{I}$ and its decomposition $\mathcal{I}^{\mathrm{active}}$, $\mathcal{I}^{\mathrm{passive}}$, and $\mathcal{I}^{\mathrm{auxiliary}}$ for a certain range of $\mathcal{E}$ (FIG.~\ref{fig:plot_flux}). As shown in FIG.~\ref{fig:plot_flux}~(a), when $\mathcal{E}<-\epsilon_l$, $\mathcal{J}_{a}^{R\to L}<0$ holds, i.e., passive transport occurs, and when $\mathcal{E}>-\epsilon_l$, $\mathcal{J}_{a}^{R\to L}>0$ holds, i.e., secondary active transport occurs. The sign of the coarse-grained information flow $\mathbb{I}$ is negative when $\mathcal{E}<-\epsilon_l$ and positive when $\mathcal{E}>-\epsilon_l$ (FIG.~\ref{fig:plot_flux}~(b)). At the same time, when $\mathcal{E}<-\epsilon_l$, $\mathcal{I}<0$ holds and the four-body correlation $\mathcal{I}^{\mathrm{active}}$ is dominant in the information flow. When $\mathcal{E}>-\epsilon_l$, $\mathcal{I}>0$ holds and the two-body correlation $\mathcal{I}^{\mathrm{passive}}$ is dominant in the information flow (FIG.~\ref{fig:plot_flux}~(c)).
Note that we plot the value of $\mathcal{I}/\sigma^{X_a}$ instead of $\mathcal{I}$ in FIG.~\ref{fig:plot_flux}~(c), but the sign of $\mathcal{I}/\sigma^{X_a}$ coincides with the sign of $\mathcal{I}$ since $\sigma^{X_a}\geq 0$ holds (see Eq.~(\ref{eq:gen_sec_law_16-state})).
Therefore, when secondary active transport occurs, $\mathbb{I}>0$, $\mathcal{I}>0$ and $\mathcal{I}\approx \mathcal{I}^{\mathrm{active}}$ hold, and when passive transport occurs, $\mathbb{I}<0$, $\mathcal{I}<0$ and $\mathcal{I}\approx \mathcal{I}^{\mathrm{passive}}$ hold. This result shows that secondary active transport and passive transport can be distinguished by the sign of $\mathbb{I}$ or $\mathcal{I}$, and the four-body correlation $\mathcal{I}^{\mathrm{active}}$ is the dominant contribution of the information flow in secondary active transport.

We also calculate the information flow for a certain range of $(\mathcal{E}, \epsilon_l, \epsilon_r)$ to see when secondary active transport occurs. As shown in FIG.~\ref{fig:heatmap}~(a), the $(\epsilon_l, \mathcal{E})$ plane is divided into three regions I ($\epsilon_l<-\mathcal{E}$), II ($\epsilon_l>\mathcal{E}$ and $\mathcal{I}>0$) and III ($\epsilon_l>\mathcal{E}$ and $\mathcal{I}<0$). In the region I and III, $\mathcal{I}<0$ holds and in the region II, $\mathcal{I}>0$ holds. This result shows that secondary active transport occurs when $\mathcal{E}>-\epsilon_l$ and $-\epsilon_l$ is sufficiently large (region I), but it does not occur even if $\mathcal{E}$ is sufficiently large when $\epsilon_l\sim 0$ (region III). Since the region II is larger for a larger value of $\epsilon_r$ as shown in FIG.~\ref{fig:heatmap}~(a), secondary active transport tends to occur when $\epsilon_r$ is large.

To see what pathways mainly occur in the regions I, II and III, we calculate the quantities $\mathcal{I}^{\mathrm{active}}$, $\mathcal{I}^{\mathrm{passive}}$ and $\mathcal{I}^{\mathrm{auxiliary}}$. As shown in FIG.~\ref{fig:heatmap}~(b), the multi-body information flow $\mathcal{I}^{\mathrm{active}}$ and $\mathcal{I}^{\mathrm{passive}}$ are dominant in the regions I and II, respectively. This behavior is consistent with the discussion in Section~\ref{subsec:sec_uni}. However, in the region III, not only the multi-body information flows $\mathcal{I}^{\mathrm{active}}$ and $\mathcal{I}^{\mathrm{passive}}$ but also the auxiliary information flow $\mathcal{I}^{\mathrm{auxiliary}}$ corresponds to a certain amount of the information flow. In other words, in the region III, the cycles 
$C^{\mathrm{active}}$ nor $C^{\mathrm{passive}}_{x_b z_b}$ do not represent the main pathway of the 16-state model.

{\red This numerical result suggests the classification of all possible spectrum of transport modes in SLC transporter proteins by the values of $\mathbb{I}$ or $\mathcal{I}$.
Though these values are difficult to measure in the experimental setting, Eqs.~(\ref{eq:second_law_X}) or (\ref{eq:second_law_16-state_X}) gives a lower bound of $\mathbb{I}$ or $\mathcal{I}$ by free energy transport $\mathcal{J}^{R\to L}_{a}(\mu^{(R)}_a-\mu^{(L)}_a)$ and binding energy $\epsilon_z$ of the transporter. Since free energy transport can be measured from experiment \cite{Adelman2016Stochastic} and free energy landscape can be obtained from the MD calculation \cite{kimanius2018uptake}, one can obtain the lower bound of $\mathbb{I}$ or $\mathcal{I}$ for actual membrane transport.

The lower bound of $\mathbb{I}$ or $\mathcal{I}$ also gives an insight for microscopic description in membrane transport. Our result shows that the four-body correlation $\mathcal{I}^{\text{active}}$ is dominant in the information flow for secondary active transport, i.e., $\mathcal{I}\approx \mathcal{I}^{\text{active}}$. Then, Eq.~(\ref{eq:second_law_16-state_X}) gives a lower bound of the four-body correlation of $X_a$, $X_b$, $Z_a$ and $Z_b$. This bound gives a partial knowledge of the probability distribution $p_s^{\text{ss}}$ of the microscopic state $s=(x_a, x_b, z_a, z_b)$ for actual membrane transport.}

\section{Conclusion}
\label{sec:conclusion}
In this work, we obtain the second law of information thermodynamics for the 16-state model representing the upper bound of the free energy transfer of the membrane transport by {\red the coarse-grained information flow $\mathbb{I}$} or the information flow $\mathcal{I}$. {\red The information flow $\mathcal{I}$ quantifies how the remaining subsystems $X_b, Z_a, Z_b$ gains information about $X_a$. From the cycle decomposition, we show the decomposition of the information flow $\mathcal{I}=\mathcal{I}^{\mathrm{active}}+\mathcal{I}^{\mathrm{passive}}+\mathcal{I}^{\mathrm{auxiliary}}$, where $\mathcal{I}^{\mathrm{active}}$ represents the four-body correlation of $X_a$, $X_b$, $Z_a$ and $Z_b$, and $\mathcal{I}^{\mathrm{passive}}$ represents the two body correlation of $X_a$ and $Z_a$ under the condition of $X_b$ and $Z_b$.} We show that the four-body correlation $\mathcal{I}^{\mathrm{active}}$ is dominant in the information flow for secondary active transport, while the two-body correlation $\mathcal{I}^{\mathrm{passive}}$ is dominant in the information flow for passive transport. This observation clarifies the role of the four-body correlation in secondary active transport; it distinguishes secondary active transport from uniport and determines the transport rate of secondary active transport.

The strategy for the cycle decomposition of the 16-state model presented in this work has high potentials to analyze the role of multi-body correlations in various biological systems composed of several subsystems. For instance, bipedal motor proteins such as kinesin-1 and myosin V and VI walk along a filament by alternating two motor domains (legs) with each repetition \cite{kinesin, myosinV, myosinVI}. To understand the relationship between the velocity of the bipedal motor proteins and the correlation between two legs, we can define a model composed of several subsystems for bipedal motor proteins and define a multi-body correlation by the cycle decomposition of this model similar to the cycle decomposition of the 16-state model presented in this work. It will be an important future problem to investigate the role of the multi-body correlation in biological systems composed of several subsystems.

{\red It would be an important future work to extend our results to the analysis of the experimental results of membrane transporters. Technical advancement of measurement would enable to estimate the coarse-grained information flow $\mathbb{I}$ or the information flow $\mathcal{I}$ by the measurement of the probability distribution $p_s^{\text{ss}}$ of the microscopic state $s=(x_a, x_b, z_a, z_b)$. Even with the current results, transient and steady-state kinetics measurements might be used to estimate the unmeasured parameters, a similar approach reported in Ref.~\cite{barreto2019transport}. In this work, the parameters for the six-state random walk model was estimated by fitting to the experimental results. Similar estimation of the parameters for the 16-state model by fitting to the experimental results would enable us to evaluate the information flow.
Alternatively, the current analysis might be extended to derive the second law of information thermodynamics given only by macroscopic quantities. Recent studies on chemical thermodynamics \cite{rao2016nonequilibrium, falasco2018information, Yoshimura2021Information, Yoshimura2021Thermodynamic} have revealed the thermodynamics inequalities of chemical reaction networks. These inequalities are closed with macroscopic quantities, namely, concentration distributions. Extending these inequalities to multi-body settings, one might obtain the analog of the information flow that can be measured in actual membrane transport.}

\begin{acknowledgements}
The authors thank M. Murao, T. Sagawa, K. Ikezaki, S. Enoki and M. Kuroda for inspiring discussion. We also thank T. Yoshii, J. Asada, T. Furuya and M. Kakiuchi for their secretarial assistance. This work was supported in part by JSPS KAKENHI grant (19H03394, 19H05794, 19H05795 to Y.O.; 20H05545 to E.M.; 19H05796, 21H01560 to S.I.), JST grant (JPMJMS2025-14, JPMJCR20E2, JPMJCR15G2, JPMJCR1852 to Y.O.; JPMJPR18M2 to S.I.), Dynamic Structural Biology program of RIKEN to Y.O., UTEC-UTokyo FSI Research Grant Program to S.I., and the Forefront Physics and Mathematics Program to Drive Transformation (FoPM) of the University of Tokyo to S.Y.

\end{acknowledgements}

\appendix

\section{Comparison of the 16-state model with the alternating access model}
\label{sec:comparison}
We compare the 16-state model with an existing unified model for secondary active transport and passive transport called the alternating access model \cite{Beckstein2021General}. The alternating access model comprises one ratchet and two types of particles, $A$ and $B$, and each ratchet has two sites. One site can bind a particle $A$, and the other can bind a particle $B$. The ratchet is in contact with two particle baths. The particle baths and the ratchet exchange particles $A$ and $B$. There are two possible forms of the ratchet: outward-facing and inward-facing. Therefore, the system has eight possible states. This model is a particular case of the 16-state model if we regard a particle in the ratchets $a$ and $b$ in the 16-state model as the particles $A$ and $B$ in the alternating access model, respectively.

{\red \section{Main reaction pathway of the 16-state model when $\mathcal{E}$ is sufficiently large}
\label{sec:main_pathway_8-state}
As discussed in subsection~\ref{subsec:sec_uni}, the 16-state model reduces to the 8-state model shown in FIG.~\ref{fig:8-state_graph}~(a) when $\mathcal{E}$ is sufficiently large. The condition (\ref{eq:assumption_for_parameter2}) leads to the inequality given by
\begin{align}
    &E_{01lr}=E_{10rl}>\max_{s\neq (0,1,l,r), (1,0,r,l)}E_s.\label{eq:energy_ineq}
\end{align}
Then, the probability that $(x_a, x_b,z_{ab})=(0,1,lr), (1,0,rl)$ is small and we can only consider the pathways composed of the remaining 6 states. Since the barrier height $\Delta_z^{(\nu)}$ satisfies Eq.~(\ref{eq:barrier_height}), we can assume that the transition of $X_a$ occurs with the particle bath $L (R)$ only when $z_a=l (r)$ and the transition of $X_b$ occurs with the particle bath $L (R)$ only when $z_b=l (r)$ for simplicity in this subsection. The pathway composed of the 6 states excluding the states $(x_a, x_b,z_{ab})=(0,1,lr), (1,0,rl)$ is the cycle $C'^{\text{active}}$ shown in  FIG.~\ref{fig:8-state_graph}~(b) or its inverse cycle. The Gibbs energy decreases in the cycle $C'^{\text{active}}$ since Eq.~(\ref{eq:assumption_for_parameter4}) is satisfied. Therefore, the main pathway is not the inverse cycle but the cycle $C'^{\text{active}}$ itself.
The pathway $C'^{\text{active}}$ is described as the cycle $C^{\text{active}}$ shown in FIG.~\ref{fig:8-state_graph}~(c) in the 16-state model.}

{\red \section{Relationship between the transport rate $\mathcal{J}^{R\to L}_{a}$ of cargo solute and the quantity $\sigma^{X_a}_{r}$}
\label{sec:relation_transport_rate}
We derive a relationship between the transport rate $\mathcal{J}^{R\to L}_{a}$ of cargo solute and the quantity $\sigma^{X_a}_{r}$ shown in Eq.~(\ref{eq:sigma_r}). The quantity $\mathcal{J}^{R\to L}_{a}$ can be written as a linear combination of $\mathbb{J}_{\tilde{e}}$ given by
\begin{align}
    \mathcal{J}^{R\to L}_{a}
    &\coloneqq \sum_{x_b, z_a, z_b}[p^{\mathrm{ss}}_{0 x_b z_a z_b}W^{A(R)}_{0\to 1| x_b z_a z_b} \nonumber\\
    &\hspace{30pt}-p^{\mathrm{ss}}_{1 x_b z_a z_b}W^{A(R)}_{1\to 0| x_b z_a z_b}]\\
    &=\sum_{z_a}\mathbb{J}_{\tilde{e}_{z_a}}^{\mathrm{ss}}
\end{align}
where $\tilde{e}_{z_a}$ is defined as $\tilde{e}_{z_a}\coloneqq ((0, z_a)\xrightarrow{R} (1, z_a))$. Since the cycle matrix $S(\tilde{e}_{z_a}, \tilde{C}_k)$ for $z_a\in \{l, r\}$ and $\tilde{C}_k\in \{\tilde{C}_1^{\tilde{X}}, \tilde{C}_2^{\tilde{X}}, \tilde{C}^{G}\}$ is given by
\begin{align}
    S(\tilde{e}_l, \tilde{C}_k)&=
    \begin{cases}
    0 & (\tilde{C}_k\in \{\tilde{C}_1^{\tilde{X}}, \tilde{C}^{G}\})\\
    1 & (\tilde{C}_k=\tilde{C}_2^{\tilde{X}})
    \end{cases},\\
    S(\tilde{e}_r, \tilde{C}_k)&=
    \begin{cases}
    1 & (\tilde{C}_k\in \{\tilde{C}_1^{\tilde{X}}, \tilde{C}^{G}\})\\
    0 & (\tilde{C}_k=\tilde{C}_2^{\tilde{X}})
    \end{cases},
\end{align}
the transport rate $\mathcal{J}^{R\to L}_{a}$ of cargo solute can be written as a linear combination of $\mathbb{J}(\tilde{C}_k)$ given by
\begin{align}
    \mathcal{J}^{R\to L}_{a}&=\mathbb{J}(\tilde{C}^{\tilde{X}}_1)+\mathbb{J}(\tilde{C}^{\tilde{X}}_2)+\mathbb{J}(\tilde{C}^{G}).
\end{align}
Therefore, $\mathcal{J}^{R\to L}_{a}$ appears in the quantity $\sigma^{X_a}_{r}$ as
\begin{align}
    &\sigma^{X_a}_{r}=\dot{\mathbb{S}}_{r}^{X_a}\nonumber\\
    &=\mathbb{J}(\tilde{C}^{\tilde{X}}_1)
    \mathbb{F}^{X_a}(\tilde{C}^{\tilde{X}}_1)
    +\mathbb{J}(\tilde{C}^{\tilde{X}}_2)
    \mathbb{F}^{X_a}(\tilde{C}^{\tilde{X}}_2)\nonumber\\
    &\hspace{12pt}+\mathbb{J}(\tilde{C}^{G})
    \mathbb{F}^{X_a}(\tilde{C}^{G})\\
    &=\mathbb{J}(\tilde{C}^{\tilde{X}}_1)
    \beta\left(
    \mu^{(R)}_a-\mu^{(L)}_a
    \right)
    +\mathbb{J}(\tilde{C}^{\tilde{X}}_2)
    \beta\left(
    \mu^{(R)}_a-\mu^{(L)}_a
    \right)\nonumber\\
    &\hspace{12pt}+\mathbb{J}(\tilde{C}^{G})
    \beta\left(
    \mu^{(R)}_a-\mu^{(L)}_a+\epsilon_{r}-\epsilon_{l}
    \right)\\
    &=\mathcal{J}^{R\to L}_{a}
    \beta\left(
    \mu^{(R)}_a-\mu^{(L)}_a
    \right)
    +\mathbb{J}(\tilde{C}^{G})
    \beta\left(
    \epsilon_{l}-\epsilon_{r}
    \right).
\end{align}
}

\section{The definition of the cycles in the 16-state model}
\label{sec:cycle_definition}
TABLE \ref{tab:cycle_definition} shows the definition of the cycles in the 16-state model. The cycles $C^{\mathrm{active}}$ and $C^{\mathrm{passive}}_{x_b z_b}$ correspond to the reaction pathways of secondary active transport and passive transport, respectively. The cycles $C^{X_a}_{x_b z_a z_b}$ and $C^{\mathrm{auxiliary}}_{x_b}$ are introduced to construct a cycle basis of the graph representing the transitions of the 16-state model.

\begin{table}[tbh]
    \begin{ruledtabular}
        \begin{tabular}{cc}
            Cycle & Definition\\
            \colrule
            $C^{\text{active}}$ & 
            \begin{tabular}{l}
            $\{(0,1,l,r)\xrightarrow{L} (1,1,l,r)\xrightarrow{L} (1,1,r,r)$\\
            $\xrightarrow{L} (1,1,r,l)\xrightarrow{L} (1,0,r,l)\xrightarrow{R} (0,0,r,l)$\\
            $\xrightarrow{L} (0,0,l,l)\xrightarrow{L} (0,0,l,r)\xrightarrow{R} (0,1,l,r)\}$
            \end{tabular}\\
            \colrule
            $C^{\text{passive}}_{x_b z_b}$ & 
            \begin{tabular}{l}
            $\{(0,x_b,l,z_b)\xrightarrow{L} (1,x_b,l,z_b)\xrightarrow{L} (1,x_b,r,z_b)$\\
            $\xrightarrow{R} (0,x_b,r,z_b) \xrightarrow{L} (0,x_b,l,z_b)\}$
            \end{tabular}\\
            \colrule
            $C^{X_a}_{x_b z_a z_b}$ & $\{(0,x_b,z_a,z_b)\xrightarrow{R}(1,x_b,z_a,z_b)\xrightarrow{L}(0,x_b,z_a,z_b)\}$\\\hline
            $C^{\text{auxiliary}}_{x_b}$ & 
            \begin{tabular}{l}
            $\{(0,x_b,l,l)\xrightarrow{L} (1,x_b,l,l)\xrightarrow{L} (1,x_b,r,l)$\\
            $\xrightarrow{L} (1,x_b,r,r)\xrightarrow{R} (0,x_b,r,r)$\\
            $\xrightarrow{L} (0,x_b,l,r)\xrightarrow{L} (0,x_b,l,l)\}$
            \end{tabular}
        \end{tabular}
    \end{ruledtabular}
    \caption{The definition the cycles used in this paper. The cycles $C^{\mathrm{active}}$ and $C^{\mathrm{passive}}_{x_b z_b}$ correspond to the reaction pathways of secondary active transport and passive transport, respectively (FIG.~\ref{fig:8-state_graph}~(c) and (d)). The cycles $C^{X_a}_{x_b z_a z_b}$ and $C^{\mathrm{auxiliary}}_{x_b}$ are introduced to construct a cycle basis of the graph representing the transitions of the 16-state model (FIG.~\ref{fig:16-state_graph}~(b) and (c)).}
    \label{tab:cycle_definition}
\end{table}

\section{The detail of the numerical analysis}
\label{appendix:numerical_analysis_detail}
We use the constants in the 16-state model as shown in TABLE \ref{tab:constants}, if not explicitly mentioned. We adopt these constants from Ref.~\cite{Muneyuki2010Allosteric}. Note that the state $z=l (r)$ is called the relaxed (energized) state in Ref.~\cite{Muneyuki2010Allosteric}. We use the different set of constants from Ref.~\cite{Muneyuki2010Allosteric}, but the constants in this paper are related to the constants in Ref.~\cite{Muneyuki2010Allosteric} by the following equations so that the transition rates $W^{(\nu)}_{s\to s'}$ for $\nu\in \{L, R\}$ and $s, s'\in \mathcal{S}$ defined in Eqs.~(\ref{eq:bipartite_16-state}) and (\ref{eq:def_W}) coincide with those in Ref.~\cite{Muneyuki2010Allosteric}. We write the subscript $\mathrm{MS}$ on the constants in Ref.~\cite{Muneyuki2010Allosteric}.

\begin{align}
    \epsilon_{z}&=\epsilon_{z, \mathrm{MS}},\\
    \Delta_{z}^{(\nu)}&=\Delta_{\nu, z, \mathrm{MS}}-10\ln {10},\\
    \ln \tau_0&=\ln \tau_{0, \mathrm{MS}} + 10\ln 10,\\
    \ln \tau_1&=\ln \tau_{0, \mathrm{MS}}+\ln \Delta^{\mathrm{AT}}_{0,0, \mathrm{MS}},\\
    \mu_{a}^{(\nu)}&=\ln [X_{II, \nu, \mathrm{MS}}],\\
    \mu_{b}^{(\nu)}&=\ln [X_{I, \nu, \mathrm{MS}}],
\end{align}
where $z\in \{l, r\}$ and $\nu\in \{L, R\}$.

\begin{table}[tbh]
    \begin{ruledtabular}
        \begin{tabular}{cccc}
             & $z_i=l$ & $z_i=r$ & \\
            \colrule
            $\epsilon_{z_i}$ & $-3\ln 10$ & $2\ln 10$ &\\
            $\Delta^{(L)}_{z_i}$ & $-2\ln 10$ & $3\ln 10$ &\\
            $\Delta^{(R)}_{z_i}$ & $0$ & $0$ &\\
            \hline\hline
             & $\nu=L$ & $\nu=R$ & \\\hline
            $\tau_0$ & \multicolumn{2}{c}{$10^{-1}$} & \\
            $\tau_1$ & \multicolumn{2}{c}{$10^{-5}$} & \\
            $\beta$ & \multicolumn{2}{c}{$1$} & \\
            $\mu^{(\nu)}_a$ & $0$ & $\ln 0.8$ & \\
            $\mu^{(\nu)}_b$ & $0$ & $\ln 0.5$ &
        \end{tabular}
    \end{ruledtabular}
    \caption{The constants shown in this table are used for the numerical analysis in Section~\ref{sec:numerical_analysis}, if not explicitly mentioned. These constants are chosen such that the transition rates $W^{(\nu)}_{s\to s'}$ for $\nu\in \{L, R\}$ and $s, s'\in \mathcal{S}$ defined in Eqs.~(\ref{eq:bipartite_16-state}) and (\ref{eq:def_W}) coincide with those in Ref.~\cite{Muneyuki2010Allosteric}.}
    \label{tab:constants}
\end{table}

\bibliography{main}

\end{document}